\documentclass[twocolumn,english,showpacs,aps,pre,secnumarabic]{revtex4}
\usepackage[T1]{fontenc}
\usepackage[latin1]{inputenc}
\usepackage{amsmath}
\usepackage{graphicx}
\usepackage{amssymb}
\usepackage{amsfonts}
\usepackage{babel}

\makeatletter

\newcommand{\noun}[1]{\textsc{#1}}
\providecommand{\tabularnewline}{\\}

\makeatother

\begin{document}

\title{Master singular behavior for the Sugden factor of the one-component
fluids near their gas-liquid critical point}

\author{Yves Garrabos{*}, Fabien Palencia, Carole Lecoutre}

\affiliation{ESEME-CNRS, ICMCB-UPR 9048, Université Bordeaux 1, 87 avenue du Docteur
Albert Schweitzer, F-33608 Pessac France.}

\email{garrabos@icmcb-bordeaux.cnrs.fr}

\author{Daniel Brosetta}

\affiliation{Laboratoire des Fluides Complexes, UMR 5150 - Université de Pau et
des Pays de l'Adour, B.P. 1155, 64013 Pau Cedex, France.}

\author{Bernard Le Neindre}

\affiliation{LIMHP-CNRS-UPR 1311, Université Paris 13, Avenue Jean Baptiste Clément,
F-93430 Villetaneuse France.}

\author{Can Erkey}

\affiliation{Department of Chemical Engineering, University of Connecticut, Storrs,
CT 06074 USA.}

\date{9 Nov. 2006}

\begin{abstract}
We present the master (i.e. unique) behavior of the squared capillary
length - so called the Sudgen factor-, as a function of the temperature-like
field along the critical isochore, asymptotically close to the gas-liquid
critical point of twenty (one component) fluids. This master behavior
is obtained using the scale dilatation of the relevant physical fields
of the one-component fluids. The scale dilatation introduces the fluid-dependent
scale factors in a manner analog with the linear relations between
physical fields and scaling fields needed by the renormalization theory
applied to the Ising-like universality class. The master behavior
for the Sudgen factor satisfies hyperscaling and can be asymptotically
fitted by the leading terms of the theoretical crossover functions
for the correlation length and the susceptibility in the homogeneous
domain recently obtained from massive renormalization in field theory.
In the absence of corresponding estimation of the theoretical crossover
functions for the interfacial tension, we define the range of the
temperature-like field where the master leading power law can be practically
used to predict the singular behavior of the Sudgen factor in conformity
with the theoretical description provided by the massive renormalization
scheme within the extended asymptotic domain of the one-component
fluid {}``subclass''. 
\end{abstract}


\maketitle

\section{Introduction}

The knowledge of interfacial properties \cite{Rowlinson 1984} for
a nonhomogeneous fluid of coexisting vapor and liquid at equilibrium
is of prime importance for many engineering applications and process
simulations. Moreover, accurate predictions of these interfacial properties
are essential to gain confidence in modeling underground geological
fluid flows in porous media, oil recovery, gas storage in geological
formations, pool boiling phenomena, microfluidic devices based on
wetting phenomena, etc. 

A large number of different forms of related phenomenological laws,
the so-called \emph{ancillary equations}, are reported in the literature
to calculate interfacial properties along the vapor-liquid equilibrium
(VLE) line \cite{Xiang 2005}. These \emph{}relations complement the
complex multiparameter equations of state (EOS's) which have been
developed to accurately fit the thermodynamic properties measured
in the homogeneous domain. Such a phenomenological approach to estimate
fluid properties is commonly based on the multiparameter corresponding-states
principle \cite{Xiang 2005,Rowlinson 1971,Ely 2000}. In the following
we call $n$-CSEOS such an EOS which contains $n\geq2$ system-dependent
parameters. The main reason for the power of such a phenomenological
approach is related to the fact that the two-parameter corresponding-states
($2$-CS) principle can be applied to any polynomial EOS which has
a liquid-vapor critical point \cite{Leach 1968}. However, in spite
of increasing the number of fluid-dependent parameters, the common
calculation of interfacial properties from ancillary equations and
$n$-CSEOS, is not only mathematically complex, but is also unable
to account for:

1) the molecular fluid complexity \cite{Rowlinson 1971}, especially
the non-spherical symmetry of molecules and the quantum behavior of
light fluids \cite{Ely 2000};

2) the \emph{asymptotic scaling of the critical phenomena} close to
the gas-liquid critical point \cite{Anisimov 2000}, especially the
non-analytic Ising-like nature \cite{Privman 1991,ZinnJustin 2002}
of the critical exponent \cite{Guida 1998}.

Among these interfacial properties, the capillary length $\ell_{Ca}$,
or more precisely the squared capillary length $\left(\ell_{Ca}\right)^{2}$
also called the Sugden factor \cite{Sudgen 1924} and noted $S_{g}$
in the following, plays a special role on Earth's gravity environment
(recalled here by the subscript $g$). The Sugden factor reflects
the balance between interfacial and volumic forces which defines the
shape and the position of the interface in equilibrium when subject
to the gravity field of constant acceleration $g$. In the case of
perfect liquid wetting,$S_{g}$ is then related to the surface tension
$\sigma$ and the density difference $\Delta\rho_{LV}=\rho_{L}-\rho_{V}$
between coexisting liquid (density $\rho_{L}$) and vapor (density
$\rho_{V}$) phases by the equation\begin{equation}
S_{g}=\frac{2\sigma}{g\Delta\rho_{LV}}\label{force balance in g field}\end{equation}
where $\sigma$ is the surface tension. Therefore, the knowledge of
the Sugden factor is an important challenge to provide better control
on non homogeneous fluid properties.

In addition, as clearly documented two decades ago \cite{Rathjen 1977 sulfurhexafluoride-halocarbons,Gielen 1984 Ar-N2-O2-CO2-CH4,Moldover 1985},
the temperature dependence of $S_{g}$, along a large temperature
range of the VLE line of all investigated one-component fluids \cite{Maass 1921 Ethane-Ethylene,Coffin 1928 i-butane,Katz 1939 2-3alcane,Stansfield 1958 Argon-Nitrogen,Smith 1967 Xenon,Grigull 1969 Carbondioxide,Gielen 1984 Ar-N2-O2-CO2-CH4,Rathjen 1977 sulfurhexafluoride-halocarbons,Straub 1980 water,Vargaftik 1983 Water,Rathjen 1980},
shows a pure \emph{power law} behavior which is applicable over an
appreciably larger temperature range {[}see below Eq. (\ref{Sg effective power law})
and the related dicussion of the Fig. \ref{fig01}a]. Such a weak
temperature dependence of the effective exponent \emph{at small but
finite} distance of the critical point was partly well-understood
to be related to the smallest value ($\simeq0.51$ \cite{Guida 1998},
see below) of the confluent exponents which govern the corrections
to asymptotic scaling of critical phenemonena \cite{Wegner 1972}.
However the theoretical reason to observe a near zero-value of the
amplitude contribution of the confluent corrections for the Sugden
factor case remains unclear, specially in the absence of estimation
of the crossover behavior of the surface tension.

Indeed, the significant theoretical improvements to account for \emph{classical-to-critical
crossover} \cite{ZinnJustin 2002}, \emph{}specially in the one-component
fluids \cite{Anisimov 2000}, provide the most powerful tools available
today to analyze accurately interfacial properties in large temperature
ranges of the non-homogeneous domain. For example in the present work,
using the crossover functions recently derived \cite{Bagnuls 2002,Garrabos 2006c}
from the massive renormalization scheme \cite{Bagnuls 1984a,Bagnuls 1984b,Bagnuls 1985,Bagnuls 1987},
our main objective is to accurately estimate this leading asymptotic
behavior of $S_{g}$ from scaling arguements \cite{Widom 1965,Fisk 1969,Stauffer 1972,Rowlinson 1984}
and available MR description \cite{Garrabos 2006a,Garrabos 2006d,Garrabos 2006e}
of the master singular behavior of the one-component fluid {}``subclass''.
Such a description is based on the formal analogy between the scale
dilatation of the physical field variables proposed by Garrabos \cite{Garrabos 1982,Garrabos 1985,Garrabos 2006b}
and the linear relation between the physical fields and the scaling
fields needed by the renormalization theory \cite{Wilson 1971}. The
major advantage of this scale dilatation method is to estimate the
universal behavior of any one-component fluids \emph{without adjustable
parameters}, by using only the four critical coordinates of its liquid-vapor
critical point (excluding here quantum fluids \cite{Garrabos 2006b}
to simplify the presentation of the scale dilatation method).

The paper is organized as follows. Section 2 demonstrates the master
singular behavior of $S_{g}$ observed from the scale dilatation method.
The corresponding Ising-like asymptotic description of $S_{g}$ based
on hyperscaling \cite{Widom 1965,Fisk 1969,Stauffer 1972,Rowlinson 1984}
and MR description \cite{Bagnuls 1984a,Bagnuls 1985,Bagnuls 1987}
of the critical crossover is reported in Section 3. The master leading
terms of the MR crossover functions for the correlation length and
the susceptibility in the homogeneous domain \cite{Bagnuls 2002,Garrabos 2006c},
are used to demonstrate that the fit of the master behavior observed
in the (nonhomogeneous) extended asymptotic domain can be made with
a theoretical precision of the same order of magnitude than the experimental
one. The discussion given in Section 4 shows the main points to be
considered for a classical-to-critical crossover description of the
interfacial properties at finite temperature distance to the critical
temperature. Specifically, we estimate precisely the temperature-like
range where this theoretical treatment becomes unappropriate to represent
the increasing non critical microscopic difference between gas and
liquid approaching the triple point temperature. Conclusion is given
in Section 5.

\section{Master singular behavior of the Sugden factor }

\subsection{The data sources}

The Sugden factor measurements $S_{g}\left(\left|\Delta T\right|\right)$
as a function of the temperature distance $\Delta T=T-T_{c}$ to the
critical point in the nonhomogeneous range $T<T_{c}$, have been published
and analyzed for several one-component fluids \cite{Maass 1921 Ethane-Ethylene,Coffin 1928 i-butane,Katz 1939 2-3alcane,Stansfield 1958 Argon-Nitrogen,Smith 1967 Xenon,Grigull 1969 Carbondioxide,Gielen 1984 Ar-N2-O2-CO2-CH4,Rathjen 1977 sulfurhexafluoride-halocarbons,Straub 1980 water,Vargaftik 1983 Water,Rathjen 1980,Moldover 1985,Grigoryev 1992 5-8alcane}.
$T$ ($T_{c}$) is the temperature (critical temperature). $S_{g}\left(\left|\Delta T\right|\right)$
data are generally obtained along the critical isochore $\rho=\rho_{c}$
in a finite temperature range bounded by max and min values of $\left|\Delta T\right|=T_{c,exp}-T$,
where $T_{c,exp}$ is the measured (or estimated) critical temperature
in the experiments {[}$\rho$ ($\rho_{c}$) is the density (critical
density)]. The relative precision clamed by the authors is generally
lower than $10\%$. For most fluids, $S_{g}$ was fitted using the
effective power law of equation \cite{Rowlinson 1984,Rathjen 1977 sulfurhexafluoride-halocarbons,Moldover 1985}
\begin{equation}
S_{g}=S_{0,e}\left|\Delta\tau^{*}\right|^{\varphi_{e}}\label{Sg effective power law}\end{equation}
where the dimensionless temperature distance $\left|\Delta\tau^{\ast}\right|$
to the critical point was defined by\begin{equation}
\left|\Delta\tau^{\ast}\right|=\frac{\left|\Delta T\right|}{T_{c,exp}}=\frac{T_{c,exp}-T}{T_{c,exp}}\label{delta tau thermal field}\end{equation}
In Eq. (\ref{Sg effective power law}), the free amplitude $S_{0,e}$
was a fluid-dependent quantity related to the effective value $\varphi_{e}\simeq0.91-0.97$
of the free (or fixed) exponent $\varphi_{e}$ considered as an adjustable
parameter when measurements were performed in a restricted temperature
range at finite distance to $T_{c,exp}$. The corresponding results
$\varphi_{e}$;$S_{0,e}$ {[}with free (or fixed) value of $\varphi_{e}$]
for each selected fluid are summarized in columns 3 and 4 of Table
\ref{tab1} (references are given in column 2). However, admitting
now that $\Delta\tau^{\ast}$ is the relevant physical field \cite{Widom 1965}
to describe the singular scaling behavior of the thermodynamic fluid
properties in the homogeneous or nonhomogenous domains along the critical
isochore, the three main {}``critical phenomena'' features of these
fitting analyses are:

i) the correlation between the effective values of $S_{0,e}$ and
$\varphi_{e}$ is highly dependent on $T_{c,exp}$ and on the (min
and max) values of the temperature range covered by the fit {}``close''
to the critical point; especially when the local values of $\varphi_{e}$
are estimated in the temperature range lower than $\left|\Delta\tau^{\ast}\right|<0.05$,
the common averaged value $\varphi_{e}=0.935\pm0.015$, equal to the
asymptotic universal value $\varphi=2\nu-\beta=0.935\pm0.015$ obtained
from the present theoretical estimation of the critical exponents
$\nu=$ and $\beta=$ \cite{Guida 1998} {[}see below Eq. (\ref{d nu phi beta})],
appears consistent with the $S_{g}$ data, whatever the one-component
fluid.

ii) accordingly, the temperature dependence of the effective exponent
over a larger temperature range is very small (then with a sign not
unambiguously defined for the small amplitude of the leading confluent
term), whatever the one-component fluid or the extension of the temperature
range of the fit;

iii) the measured value of the effective exponent is never equal to
the mean-field value $\varphi_{MF}=1$, whatever the one-component
fluid and even the large temperature range of the fit; 

As a matter of fact, it is well-established now \cite{ZinnJustin 2002}
that the range of validity of the \emph{asymptotic scaling} form of
Eq. (\ref{Sg effective power law}) is, \emph{strictly} restricted
\emph{}to the asymptotic approach of the liquid-gas critical point
(CP), when $\sigma\propto\left|\Delta\tau^{*}\right|^{\phi}$ and
$\Delta\rho_{LV}\propto\left|\Delta\tau^{*}\right|^{\beta}$ simultaneously
go to zero for $\Delta\tau^{\ast}\rightarrow0$ in Eq. (\ref{force balance in g field}).
$\phi=\left(d-1\right)\nu\approx1.261$ and $\beta\approx0.326$ are
the universal values \cite{Guida 1998} of the critical exponents
related to $\sigma$ and $\Delta\rho_{LV}$, respectively, and $\nu\approx0.63$
is the universal value \cite{Guida 1998} of the critical exponent
for the correlation length, with $\xi\propto\left|\Delta\tau^{*}\right|^{-\nu}$
(see also \cite{ZinnJustin 2002} for details of notations and definitions).
However, at small but finite $\left|\Delta\tau^{*}\right|$, any pure
power law like Eq. (\ref{Sg effective power law}) must be modified
to account for confluent corrections to scaling which can be represented
by the Wegner-like expansion \cite{Wegner 1972} with the universal
features of uniaxial 3D Ising like systems \cite{ZinnJustin 2002}.
Then, the asymptotic singular decrease of $S_{g}$ must be fitted
by the following equation\begin{equation}
S_{g}=S_{0}\left|\Delta\tau^{*}\right|^{\varphi}\left[1+{\displaystyle \sum_{i=1}^{\infty}}S_{i}\left|\Delta\tau^{*}\right|^{i\Delta}\right]\label{Wegner eq sudgen}\end{equation}
where $\Delta\approx0.51$ is the universal value \cite{Guida 1998}
of the critical exponent which characterizes the leading family of
the confluent corrections to scaling. The amplitudes $S_{0}$, $S_{1}$,
... $S_{i}$, etc., are fluid-dependent quantities. Equation (\ref{Wegner eq sudgen})
means that the critical exponent\begin{equation}
\varphi=\phi-\beta=\left(d-1\right)\nu-\beta\label{d nu phi beta}\end{equation}
only takes its universal value $\varphi\approx0.935$ asymptotically
when $\Delta\tau^{\ast}\rightarrow0$. Therefore, the weak temperature
dependence of the effective exponent at finite value of $\Delta\tau^{*}$,
first shows low rate of convergence of the Wegner expansion. Moreover,
in the fitting of the experimental $S_{g}$ data, if the contribution
of the confluent correction terms is negligible, then ${\displaystyle \sum_{i=1}^{\infty}}S_{i}\left|\Delta\tau^{*}\right|^{i\Delta}\sim0$
in Eq. (\ref{Wegner eq sudgen}).

\begin{table*}
\begin{tabular}{|c|c|c|c|cc|c|c|}
\hline 
Fluid&
Ref&
$\varphi_{e}$&
$S_{0,e}$&
$S_{0,\varphi}$&
Ref&
$\mathcal{Z}_{S,\varphi}$&
$\delta\mathcal{Z}_{S,\varphi}$\tabularnewline
&
&
&
$\left(mm^{2}\right)$&
$\left(mm^{2}\right)$&
&
&
$\%$\tabularnewline
\hline
\hline 
$Ar$&
\cite{Stansfield 1958 Argon-Nitrogen}&
$0.940$&
$4.13$&
$4.036$&
\cite{Moldover 1985}&
$2.423$&
$-1.9$\tabularnewline
&
\cite{Gielen 1984 Ar-N2-O2-CO2-CH4}&
$0.913$&
$3.78$&
$4.18$&
This work&
$2.510$&
$+1.6$\tabularnewline
\hline 
$Xe$&
\cite{Smith 1967 Xenon}&
$0.942$&
$3.05$&
$2.953$&
\cite{Moldover 1985}&
$2.668$&
$+8.0$\tabularnewline
\hline 
$N_{2}$&
\cite{Stansfield 1958 Argon-Nitrogen}&
$0.930$&
$5.46$&
$5.59$&
This work&
$2.446$&
$-1.0$\tabularnewline
&
\cite{Gielen 1984 Ar-N2-O2-CO2-CH4}&
$0.926$&
$5.10$&
$5.32$&
This work&
$2.328$&
$-5.7$\tabularnewline
\hline 
$O_{2}$&
\cite{Gielen 1984 Ar-N2-O2-CO2-CH4}&
$0.909$&
$4.85$&
$5.47$&
This work&
$2.563$&
$+4.6$\tabularnewline
\hline 
$CO_{2}$&
\cite{Grigull 1969 Carbondioxide}&
$0.933$&
$9.47$&
$9.52$&
\cite{Moldover 1985}&
$2.55$&
$+3.3$\tabularnewline
&
\cite{Gielen 1984 Ar-N2-O2-CO2-CH4}&
$0.920$&
$8.40$&
$9.0$&
This work&
$2.411$&
$-2.4$\tabularnewline
\hline 
$SF_{6}$&
\cite{Rathjen 1977 sulfurhexafluoride-halocarbons}&
$0.943$&
$3.931$&
$3.84$&
\cite{Moldover 1985}&
$2.46$&
$-0.4$\tabularnewline
\hline 
$CCl_{3}F$&
\cite{Rathjen 1977 sulfurhexafluoride-halocarbons}&
$0.928$&
$6.234$&
$6.44$&
This work&
$2.470$&
$0.0$\tabularnewline
\hline 
$CCl_{2}F_{2}$&
\cite{Rathjen 1977 sulfurhexafluoride-halocarbons}&
$0.936$&
$5.615$&
$5.589$&
This work&
$2.476$&
$+0.3$\tabularnewline
\hline 
$CClF_{3}$&
\cite{Grigull 1969 Carbondioxide}&
$0.972$&
$5.33$&
$4.5$&
This work&
$2.268$&
$-8.1$\tabularnewline
&
\cite{Rathjen 1977 sulfurhexafluoride-halocarbons}&
$0.9379$&
$4.847$&
$4.783$&
This work&
$2.411$&
$-2.4$\tabularnewline
\hline 
$CBrF_{3}$&
\cite{Rathjen 1977 sulfurhexafluoride-halocarbons}&
$0.938$&
$3.879$&
$3.826$&
This work&
$2.374$&
$-3.1$\tabularnewline
\hline 
$CHClF_{2}$&
\cite{Rathjen 1977 sulfurhexafluoride-halocarbons}&
$0.937$&
$6.859$&
$6.796$&
This work&
$2.323$&
$-6.0$\tabularnewline
\hline 
$C_{2}H_{4}$&
\cite{Maass 1921 Ethane-Ethylene}&
&
&
$13.90$&
\cite{Moldover 1985}&
$2.480$&
$+0.4$\tabularnewline
\hline 
$CH_{4}$&
\cite{Gielen 1984 Ar-N2-O2-CO2-CH4}&
$0.933$&
$13.6$&
$13.73$&
This work&
$2.382$&
$-3.6$\tabularnewline
\hline 
$C_{2}H_{6}$&
\cite{Maass 1921 Ethane-Ethylene,Katz 1939 2-3alcane}&
&
&
$14.42$&
\cite{Moldover 1985}&
$2.437$&
$-1.3$\tabularnewline
\hline 
$i-C_{4}H_{10}$&
\cite{Coffin 1928 i-butane}&
&
&
$12.71$&
\cite{Moldover 1985}&
$2.392$&
$-3.2$\tabularnewline
\hline 
$n-C_{5}H_{12}$&
\cite{Grigoryev 1992 5-8alcane}&
$0.935$&
$12.916$&
&
&
$2.488$&
$+0.7$\tabularnewline
\hline 
$n-C_{6}H_{14}$&
\cite{Grigoryev 1992 5-8alcane}&
$0.935$&
$12.753$&
&
&
$2.445$&
$-1.0$\tabularnewline
\hline 
$n-C_{7}H_{16}$&
\cite{Grigoryev 1992 5-8alcane}&
$0.935$&
$12.520$&
&
&
$2.552$&
$+3.4$\tabularnewline
\hline 
$n-C_{8}H_{18}$&
\cite{Grigoryev 1992 5-8alcane}&
$0.935$&
$12.217$&
&
&
$2.520$&
$+2.0$\tabularnewline
\hline 
$H_{2}O$&
\cite{Vargaftik 1983 Water}&
&
&
$34.72$&
\cite{Moldover 1985}&
$2.262$&
$-8.4$\tabularnewline
&
\cite{Straub 1980 water}&
$0.91$&
$33.2$&
$37.25$&
This work&
$2.427$&
$-1.7$\tabularnewline
\hline
\hline 
$\left\langle \mathcal{Z}_{S,\varphi}\right\rangle $&
&
&
&
&
&
$2.4530$&
$\pm3.1$\tabularnewline
$\mathcal{Z}_{S}$&
&
&
&
&
&
$2.47$&
\tabularnewline
\hline
\end{tabular}

\caption{Effective leading amplitude $S_{0,e}$ {[}see Eq. (\ref{Sg effective power law})]
of the Sugden factor $S_{g}\equiv\left(\ell_{Ca}\right)^{2}$ ($\ell_{Ca}$
is the capillary length) (column 2) for several one-component fluids
(colum 1); Calculated values of the master amplitude$\mathcal{Z}_{S,e}$
(column 3) of the renormalized Sugden factor $\mathcal{S}_{g^{*}}^{*}$
{[}see Eq. (\ref{master sudgen factor})], using Eq. (\ref{S0 amplitude});
The residual $\delta\mathcal{Z}_{S}=100\times\left(\frac{\left(\mathcal{Z}_{S}\right)_{exp}}{\mathcal{Z}_{S}}-1\right)$
(expressed in \%) from the value $\mathcal{Z}_{S}=2.47$ estimated
from Eq. (\ref{ZcalScal}) is given in column 4; (for data sources
and the selected fitting results see the references given in the last
column).\label{tab1}}
\end{table*}

To illustrate this behavior, the Sugden factor $S_{g}\equiv\left(\ell_{Ca}\right)^{2}$
(expressed in $m^{2}$) can be divided by $\left(T_{c}-T\right)^{\varphi}$
(expressed in $K^{\varphi}$, with $\varphi=0.935$) \cite{Moldover 1985}.
In Figure \ref{fig01}a (log-log scale), this convenient scaled form
$\frac{S_{g}}{\left(T_{c}-T\right)^{\varphi}}$ {[}$m^{2}\, K^{-\varphi}$],
is shown as a function of the temperature distance $T_{c}-T$ {[}$K$]
for eighteen one-component fluids. Each curve has a relative temperature
extension corresponding to the experimental temperature range (including
for some fluids measurements until their triple point temperature
$T_{TP}$). The use of such dimensional quantities makes the order
of magnitude of the leading amplitude contribution (i.e. $S_{g}\left(T=T_{c}+1K\right)\simeq S_{0}\left(T_{c}\right)^{-\varphi}$)
of each fluid clearly distinguishable, while the quasi-horizontal
line whatever the fluid (except the water case which needs a special
attention given in § 4.3) shows that the confluent contribution is
negligible (i.e. ${\displaystyle \sum_{i=1}^{\infty}}S_{i}\left(T_{c}\right)^{-i\Delta}\left(T_{c}-T\right)^{i\Delta}\sim0$).
$S_{g}$ values at $T_{c}-T=1\, K$ cover one decade: from $1.5\,10^{-8}\, m^{2}$
for sulfurhexafluoride {[}with $S_{0}\left(SF_{6}\right)=3.84\, mm^{2}$],
to $1.5\,10^{-7}\, m^{2}$ for methane {[}$S_{0}\left(CH_{4}\right)=13.80\, mm^{2}$].

We have noted that some of the data reported in Fig. \ref{fig01}a
have been measured in a large temperature range of the coexisting
VLE line, including measurements close to $T_{TP}$. To separate the
asymptotic critical range from the triple point location along the
temperature axis, we have marked by vertical arrows the temperature
distance where $T=0.7\, T_{c}$ (i.e. the temperature distance where
the practical fluid-dependent acentric factor \cite{Pitzer1955} is
defined in the nonhomogeneous domain). The {}``non-critical'' temperature
range between $0.3\, T_{c}\leq T_{c}-T\leq T_{c}-T_{TP}$ (the right
hand side of the corresponding arrows in Fig. \ref{fig01}a) is considered
to be far away from the critical point. Close to the critical point,
the practical temperature range where the Wegner-like expansion fit
the singular behavior does not usually exceed a few percent in \textbf{$\left|\Delta\tau^{*}\right|$}
\cite{Levelt-Sengers 1981}. In a similar arbitrary manner, we have
represented by vertical arrows the temperature distance where $T=0.99\, T_{c}$,
to make clear the {}``critical'' temperature range $T_{c}-T\leq0.01\, T_{c}$
(the left hand side of the corresponding arrows in Fig. \ref{fig01}a),
where the use of Eq. (\ref{Wegner eq sudgen}) has a theoretical justification,
as discussed below in § 3. To introduce the main physical parameters
needed for accurate description of the singular behavior of $S_{g}$
in this asymptotic temperature range, the next subsection presents
the application of the scale dilatation method \cite{Garrabos 1982,Garrabos 1985}
leading to define the dimensionless form (noted $S_{g}^{*}$) and
the renormalized form (noted $\mathcal{S}_{g^{*}}^{*}$) of the Sugden
factor, with two objectives:

1) to show that any modeling based on the $2$-CS principle is inaccurate
to describe the fluid dependence of \emph{}(dimensionless) $S_{g}^{*}$
{[}see below Eq. (\ref{Sstarg})] as a function of the (dimensionless)
temperature field $\Delta\tau^{*}$;

2) to unambiguously show the master singular behavior of \emph{}(renormalized)
$\mathcal{S}_{g^{*}}^{*}$ {[}see below Eq. (\ref{master sudgen factor})]
as a function of the (renormalized) temperature-like field, noted
$\mathcal{T}^{*}$ {[}see below Eq. (\ref{delta tau dilatation})]
.

\subsection{The scale dilatation method to observe the master singular behavior}

The following analysis of the Sugden factor from the scale dilatation
method is similar to the one of the correlation length given in Ref.
\cite{Garrabos 2006a}. We recall only the main features (ignoring
the quantum contributions at $T\cong T_{c}$ \cite{Garrabos 2006b}).
The input data are the four critical coordinates\begin{equation}
Q_{c,a_{\bar{p}}}^{min}=\left\{ T_{c},v_{\bar{p},c},p_{c},\gamma_{c}^{\prime}=\left[\left(\frac{\partial p}{\partial T}\right)_{v_{\bar{p},c}}\right]_{CP}\right\} \label{qcmin critical coordinates}\end{equation}
which localize the liquid gas critical point on the phase surface
of equation $\Phi_{a_{\bar{p}}}^{p}\left(p,v_{\bar{p}},T\right)=0$
for each fluid particle of mass $m_{\bar{p}}$ \cite{particle mass}.
$p$ is the pressure, $v_{\bar{p}}$ is the particle volume, and $a_{\bar{p}}\left(T,v_{\bar{p}}\right)$
is the Helmholtz energy per particle. The subscript $\bar{p}$ refers
to a particle quantity and all the definitions and notations related
to Eq. (\ref{qcmin critical coordinates}) are given in \cite{Garrabos 1982,Garrabos 1985,Garrabos 2006b}.
The critical data related to the fluids selected in Table \ref{tab1}
are reported in Table \ref{tab2}. We note that $T_{c}$ values of
Table \ref{tab2}, which were obtained from the thermodynamic analysis
of the phase surface, can be slightly different from $T_{c,exp}$
values given in the experiments refered in Table \ref{tab1}. Also
$\rho_{c}=\frac{m_{\bar{p}}}{v_{\bar{p},c}}$ values from Table \ref{tab2}
can be slightly different from the experimental critical density values
reported in these experiments.

\begin{table*}
\begin{tabular}{|c|c|c|c|c|c|c|c|c|c|}
\cline{2-2} \cline{3-3} \cline{4-4} \cline{5-5} \cline{6-6} \cline{7-7} \cline{8-8} \cline{9-9} \cline{10-10} 
\multicolumn{1}{|c|}{Fluid}&
 $\begin{array}{c}
m_{\bar{p}}\\
\left(10^{-26}kg\right)\end{array}$&
 $\begin{array}{c}
T_{c}\\
\left(K\right)\end{array}$&
 $\begin{array}{c}
v_{\bar{p},c}\\
\left(nm^{3}\right)\end{array}$&
 $\begin{array}{c}
p_{c}\\
\left(MPa\right)\end{array}$&
 $\begin{array}{c}
\gamma_{c}^{\prime}\\
\left(MPa\, K^{-1}\right)\end{array}$&
 $\begin{array}{c}
\left(\beta_{c}\right)^{-1}\\
\left(10^{-21}J\right)\end{array}$&
 $\begin{array}{c}
\alpha_{c}\\
\left(nm\right)\end{array}$&
 $Z_{c}$&
 $Y_{c}$\tabularnewline
\hline
\multicolumn{1}{|c|}{$Ar$}&
 $6.634$&
 $150.725$&
 $0.12388$&
 $4.865$&
 $0.191$&
 $2.08099$&
 $0.753463$&
 $0.2896$&
 $4.32882$\tabularnewline
\hline
\multicolumn{1}{|c|}{$Xe$}&
 $21.803$&
 $289.733$&
 $0.19589$&
 $5.84$&
 $0.1182$&
 $4.0003$&
 $0.881508$&
 $0.28601$&
 $4.85434$\tabularnewline
\hline
\multicolumn{1}{|c|}{$N_{2}$}&
$4.652$&
$126.214$&
$0.14814$&
$3.398$&
$0.1715$&
$1.74258$&
$0.80043$&
$0.288875$&
$5.37014$\tabularnewline
\hline
\multicolumn{1}{|c|}{$O_{2}$}&
$5.314$&
$154.580$&
$0.12187$&
$5.043$&
$0.1953$&
$2.13421$&
$0.750786$&
$0.287972$&
$4.98641$\tabularnewline
\hline
\multicolumn{1}{|c|}{$CO_{2}$}&
 $7.308$&
 $304.137$&
 $0.15622$&
 $7.3753$&
 $0.170$&
 $4.19907$&
 $0.82882$&
 $0.27438$&
 $6.0104$\tabularnewline
\hline
\multicolumn{1}{|c|}{$SF_{6}$}&
 $24.255$&
 $318.735$&
 $0.32769$&
 $3.754$&
 $0.0835$&
 $4.40062$&
 $1.0544$&
 $0.27954$&
 $6.0896$\tabularnewline
\hline
\multicolumn{1}{|c|}{$CCl_{3}F$}&
$22.810$&
$471.110$&
$0.41174$&
$4.4076$&
$0.0655$&
$6.50438$&
$1.13850$&
$0.27901$&
$6.00530$\tabularnewline
\hline
\multicolumn{1}{|c|}{$CCl_{2}F_{2}$}&
$20.078$&
$384.930$&
$0.35562$&
$4.1249$&
$0.0745$&
$5.31454$&
$1.08814$&
$0.27602$&
$5.95186$\tabularnewline
\hline
\multicolumn{1}{|c|}{$CClF_{3}$}&
$17.348$&
$301.88$&
$0.29807$&
$3.877$&
$0.0910$&
$4.16791$&
$1.02441$&
$0.27727$&
$6.08861$\tabularnewline
\hline
\multicolumn{1}{|c|}{$CBrF_{3}$}&
$24.727$&
$340.19$&
$0.33191$&
$3.956$&
$0.0810$&
$4.69683$&
$1.05889$&
$0.27956$&
$5.96985$\tabularnewline
\hline
\multicolumn{1}{|c|}{$CHClF_{2}$}&
$14.359$&
$369.30$&
$0.27454$&
$4.990$&
$0.0965$&
$5.09874$&
$1.00721$&
$0.26869$&
$6.14259$\tabularnewline
\hline
\multicolumn{1}{|c|}{$C_{2}H_{4}$}&
$4.658$&
$282.345$&
$0.21667$&
$5.042$&
$0.11337$&
$3.89820$&
$0.91781$&
$0.28131$&
$5.34856$\tabularnewline
\hline
\multicolumn{1}{|c|}{$CH_{4}$}&
$2.664$&
$190.564$&
$0.16361$&
$4.5992$&
$0.14442$&
$2.63102$&
$0.830133$&
$0.28679$&
$4.9838$\tabularnewline
\hline
\multicolumn{1}{|c|}{$C_{2}H_{6}$}&
$4.993$&
$305.322$&
$0.24171$&
$4.872$&
$0.10304$&
$4.21554$&
$0.95290$&
$0.27935$&
$5.45505$\tabularnewline
\hline
\multicolumn{1}{|c|}{$i-C_{4}H_{10}$}&
$9.652$&
$407.844$&
0.43020&
$3.629$&
$0.0610$&
$5.63084$&
$1.15770$&
$0.27726$&
$5.93407$\tabularnewline
\hline
\multicolumn{1}{|c|}{$n-C_{5}H_{12}$}&
$11.981$&
$469.70$&
$0.521785$&
$3.3665$&
$0.0511$&
$6.48491$&
$1.24425$&
$0.270875$&
$6.12956$\tabularnewline
\hline
\multicolumn{1}{|c|}{$n-C_{6}H_{14}$}&
$14.310$&
$507.49$&
$0.61138$&
$3.0181$&
$0.043658*$&
$7.00666$&
$1.3186$&
$0.26667$&
$6.30719$\tabularnewline
\hline
\multicolumn{1}{|c|}{$n-C_{7}H_{16}$}&
$16.6386$&
$540.13$&
$0.7168$&
$2.727$&
$0.038068*$&
$7.45731$&
$1.3983$&
$0.26218$&
$6.64356$\tabularnewline
\hline
\multicolumn{1}{|c|}{$n-C_{8}H_{18}$}&
$18.9683$&
$568.88$&
$0.81839$&
$2.486$&
$0.033768*$&
$7.85424$&
$1.46746$&
$0.258978$&
$6.82776$\tabularnewline
\hline
$H_{2}O$&
$2.991$&
$647.067$&
$0.09268$&
$22.046$&
$0.275$&
$8.93373$&
$0.740$&
$0.229$&
$7.071$\tabularnewline
\hline
\end{tabular}

\caption{Set of critical parameters {[}see Eqs. (\ref{qcmin critical coordinates})
and (\ref{qcmin scale factors})] for twenty one-component fluids
of particle mass $m_{\bar{p}}$ \cite{particle mass} selected in
the present work (see Fig. 1).\label{tab2}}
\end{table*}

In combining $Q_{c,a_{\bar{p}}}^{\min}$, the Boltzmann constant $k_{B}$,
and $d=3$, Eq. (\ref{qcmin critical coordinates}) can be written
in a more convenient form, such that\begin{equation}
Q_{c}^{\min}=\left\{ \left(\beta_{c}\right)^{-1},\alpha_{c},Z_{c},Y_{c}\right\} \label{qcmin scale factors}\end{equation}
which introduces the following four scale factors given by, \begin{equation}
\left(\beta_{c}\right)^{-1}=k_{B}T_{c}\label{energy scale}\end{equation}
\begin{equation}
\alpha_{c}=\left(\frac{k_{B}T_{c}}{p_{c}}\right)^{\frac{1}{d}}\label{length scale}\end{equation}
\begin{equation}
Z_{c}=\frac{p_{c}v_{\bar{p},c}}{k_{B}T_{c}}\label{critical compression factor}\end{equation}
 \begin{equation}
Y_{c}=\gamma_{c}^{\prime}\frac{T_{c}}{p_{c}}-1\label{critical isochoric factor}\end{equation}
 Equation (\ref{qcmin scale factors}) involves one energy scale unit
$\left(\beta_{c}\right)^{-1}$, one length scale unit $\alpha_{c}$,
and two dimensionless \emph{scale factors} $Z_{c}$ and $Y_{c}$ characterizing
two preferred directions to cross the critical point along the critical
isotherm and the critical isochore, respectively. $\alpha_{c}$, which
does not depend of the size $L\sim\left(V\right)^{\frac{1}{d}}$ of
the container, has a clear physical meaning as length unit \cite{Garrabos 1982}:
it represents the spatial extent of the short-ranged (Lennard-Jones
like) molecular interaction \cite{Hirschfelder 1964}, which allows
us to define $v_{c,I}=\frac{k_{B}T_{c}}{p_{c}}$ as the volume of
the \textit{\emph{microscopic}} \textit{critical interaction cell}
(CIC) of each fluid. $Z_{c}$ is the usual critical compression factor.
Furthermore, $\left(Z_{c}\right)^{-1}=n_{c}v_{c,I}$ is the number
of particles that fills $v_{c,I}$. \emph{}Then the minimal set of
data in Eq. (\ref{qcmin scale factors}) \emph{}is related to the
thermodynamic properties of the \emph{}critical interaction cell \emph{}of
size $\alpha_{c}=\left(v_{c,I}\right)^{\frac{1}{d}}$ \cite{Garrabos 2005}.

We recall that the critical compression factor $Z_{c}$, and the critical
Riedel factor $\alpha_{R,c}$ \cite{Riedel 1954} {[}related to $Y_{c}$
by $\alpha_{R,c}=Y_{c}+1$], are among the basic parameters used to
develop $4$-CSEOS's for engineering fluid modeling \cite{Poling 2001}.

\begin{figure*}
\includegraphics[width=178mm,keepaspectratio]{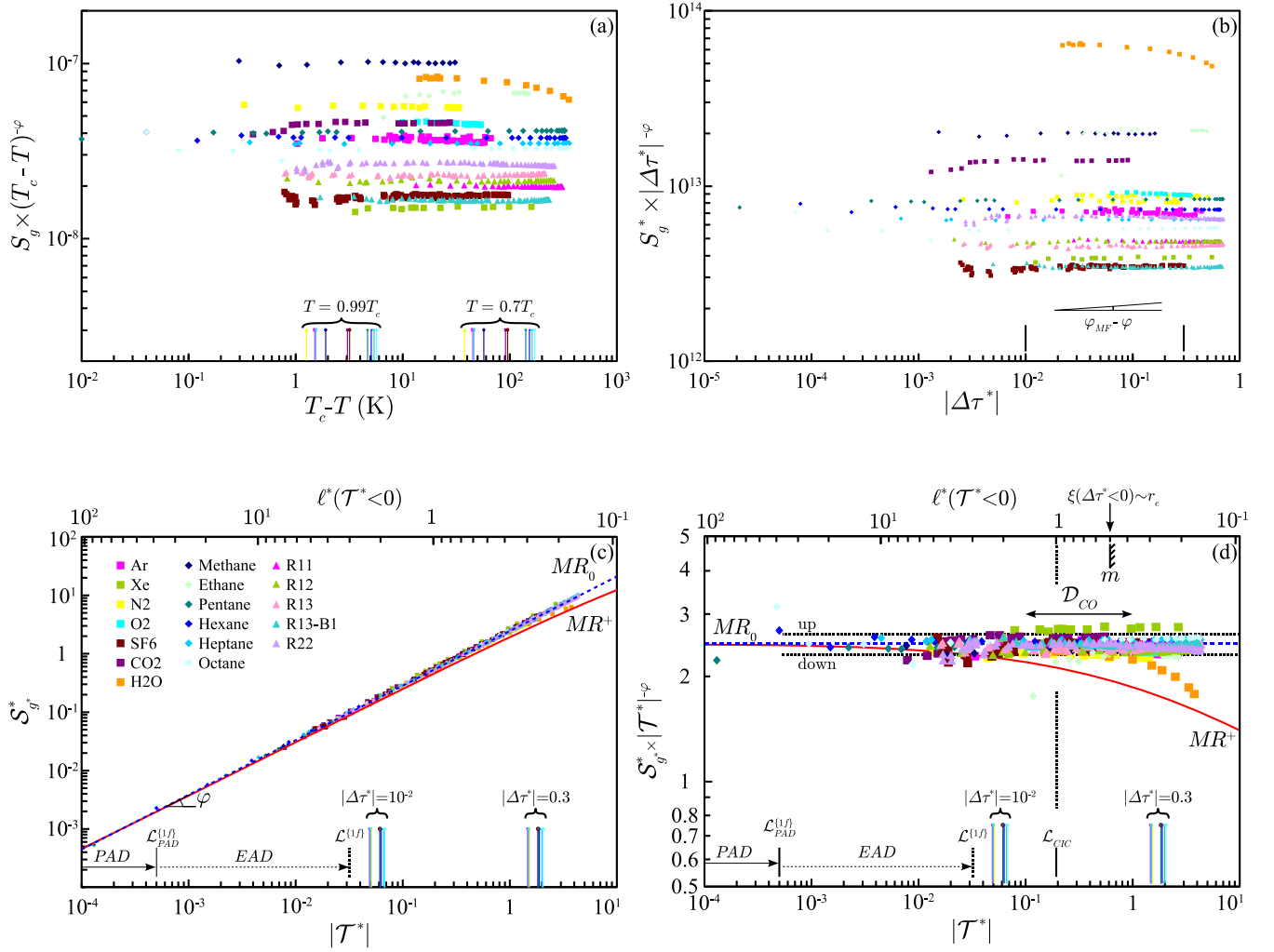}

\caption{(Color online) a) Singular behavior (log-log scale) of $\frac{S_{g}}{\left(T_{c}-T\right)^{\varphi}}$
(expressed in $m^{2}\, K^{-\varphi}$, with $\varphi=0.935$), as
a function of the temperature distance $T_{c}-T$, for nonhomogeneous
one-component fluids (see Tables I and II). For each fluid, the arrows
indicate the arbitrary temperature distances of $T_{c}-T=0.01T_{c}$
(left) and $T_{c}-T=0.3T_{c}$ (rigth); Inserted Table gives the color
indexation for each fluid (see Table I and text for details); b) Log-Log
scale of the respective dimensionless variables using $\alpha_{c}$
as a length unit and $\left(\beta_{c}\right)^{-1}$ as a energy unit.
Each distinguishable curve illustrates the failure of the classical
corresponding state scheme. The expected slopes of a \emph{classical}
power law with mean field exponent $\varphi_{MF}$ are given by the
directions labelled $MF$ (see text); c) Master singular behavior
(log-log scale) of the renormalized Sugden factor $\mathcal{S}_{g^{*}}^{*}=g^{*}\left(Z_{c}\right)^{-\frac{3}{2}}S_{g}^{*}$
(with $S_{g}^{*}=\frac{S_{g}}{\left(\alpha_{c}\right)^{2}}$ and $g^{*}=m_{\bar{p}}\beta_{c}\alpha_{c}g$),
as a function of the renormalized thermal field $\mathcal{T}^{*}=Y_{c}\left|\Delta\tau^{*}\right|$
{[}see text and Eq. (\ref{scale dilatation})]; d) Master {}``confluent''
behavior of the rescaled quantity $\frac{\mathcal{S}_{g^{*}}^{*}}{\left|\mathcal{T}^{*}\right|^{\varphi}}$,
as a function of the renormalized \emph{}thermal field $\mathcal{T}^{*}=Y_{c}\left|\Delta\tau^{*}\right|$,
{[}see text and Eq. (\ref{scale dilatation})]. In c) and d): The
curve labelled $MR^{+}$ corresponds to the Eq. (\ref{master Scalhat vs master khitstar elestar})
using crossover Eqs. (\ref{elemaster complete equation}) and (\ref{khimaster complete equation})
for the correlation length and the susceptibility, respectively, in
the homogeneous domain; $\mathcal{L}_{PAD}^{\left\{ 1f\right\} }$
{[}Eq. (\ref{PAD1f extension})] and $\mathcal{L}^{\left\{ 1f\right\} }$
{[}see Eq. (\ref{EAD1f extension})] correspond to the extension of
the preasymptotic domain and the extended asymptotic domain, respectively;
The graduation of the upper horizontal axis gives $\frac{\ell^{*}\left(\mathcal{T}^{*}>0\right)}{1.96}$
calculated from master crossover Eq. (\ref{elemaster complete equation});
At large values of the renormalized thermal field ($\mathcal{T}^{*}\geq0.2$)
which correspond to $\frac{\ell^{*}\left(\mathcal{T}^{*}>0\right)}{1.96}\lesssim1$,
the restricted range labelled $\mathcal{D}_{EC}$ indicates the effective
crossover for the exponent $\beta$, while the vertical index labelled
$m$ gives a rough estimate of the \emph{microscopic} limit $\frac{\ell^{*}\left(\mathcal{T}^{*}>0\right)}{1.96}\approx\frac{1}{2}$
where the correlation length $\xi$ in the nonhomogeneous domain will
reach the order of magnitude of the two particle equilibrium position
$r_{e}$ (with $r_{e}\gtrapprox\sigma$, where $\sigma$ is the \emph{size}
of the particle); See inserted Table and legend of a) for fluid color
indexation; See legend of b) for arrows in lower horizontal axis.\label{fig01}}
\end{figure*}

The characteristic units $\left(\beta_{c}\right)^{-1}$ and $\alpha_{c}$
are the parameters needed to provide a dimensionless analysis of the
fluid properties, leading to their {}``classical'' description based
on the two-parameter corresponding state ($2$-CS) description. Obviously,
the dimensionless form of the Sugden factor is given by \begin{equation}
S_{g}^{*}=\frac{S_{g}}{\left(\alpha_{c}\right)^{2}}\label{Sstarg}\end{equation}
Figure \ref{fig01}b (log-log scale; color online) represents the
confluent behavior of the rescaled dimensionless quantity $\frac{S_{g}^{*}}{\left|\Delta\tau^{*}\right|^{\varphi}}$
as a function of the dimensionless temperature distance $\Delta\tau^{\ast}$.
Figure \ref{fig01}b complements Fig. 3 initially published by Moldover
in Ref. \cite{Moldover 1985}, after normalization of the vertical
axis by $\left(\alpha_{c}\right)^{2}$. Figure \ref{fig01}b illustrates
the results of any classical two-parameter corresponding state theory
(here the two characteristic parameters are $\left(\beta_{c}\right)^{-1}$
and $\alpha_{c}$). Figure \ref{fig01}b, shows the failure of the
$2$-CS principle in terms of molecular fluid complexity since, from
xenon to water, the dimensionless Sugden factor covers one order of
magnitude at the same reduced temperature distance to the critical
point. Moreover, in terms of \emph{classical} critical phenomena,
using Eq. (\ref{force balance in g field}) where $\Delta\rho_{LV}\propto\left|\Delta\tau^{*}\right|^{\beta_{MF}}$
and $\sigma\propto\left|\Delta\tau^{*}\right|^{\phi_{MF}}$ with $\beta_{MF}=\frac{1}{2}$
and $\phi_{MF}=\frac{3}{2}$ \cite{Widom 1996}, we obtain the \emph{mean
field} exponent $\varphi_{MF}=1$. This mean-field value associated
to the \textit{classical} behavior of the correlation length (with
exponent $\nu_{MF}=\frac{1}{2}$) expected from Van der Waals-like
theories \cite{Widom 1996,Kostrowicka 2004}, is unable to describe
the experimental results, even at large temperature distance, as shown
by the significantly positive slope $\varphi_{MF}-\varphi\simeq0.065$
reported in Fig. \ref{fig01}b. In addition, the scaling law $\left(d-1\right)\nu=\phi$,
that explicitly involve $d$, is not correct for mean-field exponents
in three dimension. We will turn back on the mean-field theories in
§ 4.2 when we will discuss the related critical-to-critical crossover
description of the interfacial properties.

In the next step, the dimensionless scale factors $Y_{c}$ and $Z_{c}$
are introduced throughout the scale dilatation method \cite{Garrabos 1985}.
Typically, the scale dilatation of the dimensionless temperature distance,
\begin{equation}
\Delta\tau^{\ast}=k_{B}\beta_{c}\left(T-T_{c}\right)\label{thermal field}\end{equation}
 leads to the \emph{}renormalized \emph{}thermal field,

\begin{equation}
\mathcal{T}^{\ast}=Y_{c}\Delta\tau^{*}\label{delta tau dilatation}\end{equation}
The scale dilatation of the dimensionless order parameter density\begin{equation}
\Delta m^{*}=\left(\alpha_{c}\right)^{d}\left(n-n_{c}\right)=\left(Z_{c}\right)^{-1}\Delta\tilde{\rho}\label{reduced number density distance}\end{equation}
leads to the renormalized order parameter density\begin{equation}
\mathcal{M}^{*}=\left(Z_{c}\right)^{\frac{d}{2}}\Delta m^{*}=\left(Z_{c}\right)^{\frac{1}{2}}\Delta\tilde{\rho}\label{master order parameter}\end{equation}
In addition, the renormalized form $\ell^{*}\equiv\xi^{*}=\frac{\xi}{\alpha_{c}}$
of the correlation length $\xi$ \cite{Garrabos 2006b}, leads to
the \emph{}renormalized form, noted $\Sigma^{*}$, of the surface
tension $\sigma$ such that \cite{LeNeindre 2002}\begin{equation}
\Sigma^{*}\equiv\sigma^{*}=\sigma\beta_{c}\left(\alpha_{c}\right)^{d-1}\label{master surface tension}\end{equation}
 Taking into account Eqs. (\ref{force balance in g field}) and (\ref{Sstarg}),
the renormalized Sugden factor $\mathcal{S}_{g^{*}}^{*}$ reads as
follows \cite{LeNeindre 2002}\begin{equation}
\mathcal{S}_{g^{*}}^{*}=g^{*}\left(Z_{c}\right)^{-\frac{3}{2}}\left(\ell_{Ca}^{*}\right)^{d-1}=g^{*}\left(Z_{c}\right)^{-\frac{3}{2}}S_{g}^{*}\label{master sudgen factor}\end{equation}
with $g^{*}=m_{\bar{p}}\beta_{c}\alpha_{c}g$. Therefore, after application
of the scale dilatation method, the renormalized form of Eq. (\ref{force balance in g field})
is given by \begin{equation}
\mathcal{S}_{g^{*}}^{*}=\frac{\Sigma^{*}}{\mathcal{M}_{LV}^{*}}\label{master balance eq}\end{equation}
As expected \cite{Garrabos 1982}, the collapse on the master curve
obtained from the scale transformations\begin{equation}
\begin{array}{cl}
\Delta\tau^{*} & \rightarrow\mathcal{T}^{\ast}=Y_{c}\Delta\tau^{*}\\
S_{g}^{*} & \rightarrow\mathcal{S}_{g^{*}}^{*}=g^{*}\left(Z_{c}\right)^{-\frac{3}{2}}S_{g}^{*}\;\;\left(T1\, case\right)\\
 & \rightarrow\mathcal{S}_{g^{*}}^{*}\times\left|\mathcal{T}^{\ast}\right|^{-\varphi}\;\;\;\;\;\;\;\;\;\;\;\left(T2\, case\right)\end{array}\label{scale dilatation}\end{equation}
is shown in Fig. \ref{fig01}c ($T1\, case$) and \ref{fig01}d ($T2\, case$),
\emph{independently of any theoretical form used to represent this
master behavior}. Now the scatter of the collapsed data corresponds
to the estimated precision ($10\%$) for the Sugden factor of each
fluid.

\subsection{Predictive power of the scale dilatation method within the Ising-like
preasymptotic domain}

As initially shown in Ref. \cite{Garrabos 1985}, we can expect to
fit the master singular behavior of $\mathcal{S}_{g^{*}}^{*}$ observed
asymptotically close to the critical temperature by a \emph{restricted}
(two-term) Wegner-like expansion given by\begin{equation}
\mathcal{S}_{g^{*}}^{*}=\mathcal{Z}_{\mathcal{S}}\left|\mathcal{T}^{*}\right|^{\phi}\left[1+\mathcal{Z}_{\mathcal{S}}^{1}\left|\mathcal{T}^{*}\right|^{\Delta}\right]\label{two-term master Sgstar}\end{equation}
where $\phi\approx0.935$ and $\Delta\approx0.51$ are the universal
critical exponents while $\mathcal{Z}_{\mathcal{S}}$ and $\mathcal{Z}_{\mathcal{S}}^{1}$
are the master (i.e. unique) leading and confluent amplitudes, respectively,
for all one-components fluids. By term to term comparison of Eqs.
(\ref{Wegner eq sudgen}) and (\ref{two-term master Sgstar}) using
Eqs. (\ref{scale dilatation}), we obtain the following relations\begin{equation}
\mathcal{Z}_{\mathcal{S}}=g^{*}\left(\alpha_{c}\right)^{1-d}\left(Z_{c}\right)^{-\frac{3}{2}}\left(Y_{c}\right)^{-\varphi}S_{0}\label{S0 amplitude}\end{equation}
\begin{equation}
\mathcal{Z}_{\mathcal{S}}^{1}=\left(Y_{c}\right)^{-\Delta}S_{1}\label{S1 amplitude}\end{equation}
which shows the unequivocal link between master amplitudes and system-dependent
amplitudes, through $Q_{c}^{\min}$ {[}(Eq. (\ref{qcmin scale factors})].

In other words, only when the fluid-dependent set $Q_{c}^{\min}$
and the master amplitudes $\mathcal{Z}_{\mathcal{S}}$ and $\mathcal{Z}_{\mathcal{S}}^{1}$
are known, the restricted Wegner-like expansion {[}Eq. (\ref{Wegner eq sudgen})
with $i\leq1$] of $S_{g}$ can be determined for any one-component
fluid by inverting Eqs.(\ref{S0 amplitude}) and (\ref{S1 amplitude}),
such that $S_{0}=\left(\alpha_{c}\right)^{d-1}\left(g^{*}\right)^{-1}\left(Z_{c}\right)^{\frac{3}{2}}\left(Y_{c}\right)^{\varphi}\mathcal{Z}_{\mathcal{S}}$
and $S_{1}=\left(Y_{c}\right)^{\Delta}\mathcal{Z}_{\mathcal{S}}^{1}$.
Then, the master values of $\mathcal{Z}_{\mathcal{S}}$ and $\mathcal{Z}_{\mathcal{S}}^{1}$
conform to the universal features calculated for the Ising-like universality
class (i.e., some combinations and ratios of $\mathcal{Z}_{\mathcal{S}}$
and $\mathcal{Z}_{\mathcal{S}}^{1}$ take universal values, in agreement
with the two-scale-factor universality). We will detail this point
in § 4. Before, in the next Section, the scale transformations of
Eq. (\ref{scale dilatation}) are reported in conformity with the
asymptotic linearization \cite{Wilson 1971} of the two relevant fields
needed by the renormalization group theory. That leads indeed to the
correct account for universal features estimated in the preasymptotic
domain and the accurate determination of $\mathcal{Z}_{\mathcal{S}}$
using the present theoretical status provided by the MR scheme \cite{Guida 1998,Bagnuls 2002}.

\section{Ising-like crossover functions for the Sugden factor}

To our knowledge, the theoretical function giving the classical-to-critical
crossover of the interfacial tension $\sigma\left(\Delta\tau^{*}\right)$
is not available from the MR scheme, while the one of the coexisting
density $\Delta\rho_{LV}\left(\Delta\tau^{*}\right)$ \cite{Garrabos 2006c}
remains affected by a large uncertainty on the value of the first
confluent amplitude. Therefore, using either Eq. (\ref{force balance in g field})
for physical properties or Eq. (\ref{master balance eq}) for renormalized
properties, the related crossover functions of the physical and renormalized
Sugden factor remain undetermined. Especially the value of $\mathcal{Z}_{\mathcal{S}}^{1}$
{[}$S_{1}$, respectively] in Eq. (\ref{two-term master Sgstar})
{[}Eq. (\ref{Wegner eq sudgen}), respectively] cannot be estimated
from theoretical prediction of the universal value of the confluent
amplitude ratios related to the lowest confluent exponent $\Delta$
(see also our discussion in Section 4.1). However, \emph{hyperscaling}
related to the two-scale-factor universality of the asymptotic Ising-like
description provides unambiguous determination of the values of $\mathcal{Z}_{\mathcal{S}}$
{[}$S_{0}$, respectively] in Eq. (\ref{two-term master Sgstar})
{[}Eq. (\ref{Wegner eq sudgen}), respectively]. This determination
is presented below using the master forms of Ising-like crossover
functions obtained from the massive renormalization (MR) scheme. 

However, we note that a form equivalent to Eq. (\ref{two-term master Sgstar})
was also recovered in the crossover approach of Belyakov et al \cite{Belyakov 1995},
who uses adjustable parameters as scale factors of the physical variables.
The solution was obtained on the basis of the $\epsilon$-expansion
in first order $\epsilon$ and was not considered here due to the
arbitrary of the phenomenological adjustement to provide the crossover
to a classical behavior.

\subsection{Asymptotic hyperscaling description of the Sugden factor}

It is well-established experimentally \cite{Gielen 1984 Ar-N2-O2-CO2-CH4,Moldover 1985}
and theoretically \cite{Rowlinson 1984,Mon 1988,Shaw 1989,Fisher 1998},
that the asymptotic limit for $\Delta\tau^{*}\rightarrow0$ of the
product of the interfacial tension by the squared correlation length
takes a universal value, noted $R_{\sigma\xi}^{\pm}$, for the Ising-like
universality class. This result proceeds from the Widom's scaling
law between the corresponding critical exponents $\nu$ and $\mu$
given by \begin{equation}
\left(d-1\right)\nu=\phi\label{Widom scaling law}\end{equation}
 with $d=3$ in our present study. Therefore, we can introduce $R_{\sigma\xi}^{\pm}$
as follows\begin{equation}
R_{\sigma\xi}^{\pm}=\beta_{c}\times\lim\left[\sigma\left(\left|\Delta\tau^{*}\right|\right)\times\left[\xi\left(\Delta\tau^{*}\right)\right]^{d-1}\right]_{\Delta\tau^{*}\rightarrow0^{\pm}}\label{Rsigmaksi}\end{equation}
 where the superscript $\pm$ refers to the singular behavior of $\xi$
above ($+$) or below ($-$) $T_{c}$. As a matter of fact, accounting
for the universal ratio $\frac{\xi\left(\Delta\tau^{*}>0\right)}{\xi\left(\Delta\tau^{*}<0\right)}=1.96$
for the Ising-like universality class \cite{ZinnJustin 2002}, the
amplitude combination $R_{\sigma\xi}^{+}=\left(1.96\right)^{d-1}R_{\sigma\xi}^{-}$
shows that an interfacial property (here $\sigma\propto\left|\Delta\tau^{*}\right|^{\phi}$)
in the non-homogeneous domain ($\Delta\tau^{*}<0$) is related in
a universal manner to the correlation length in the homogeneous domain
($\Delta\tau^{*}>0$).

Considering the scaling law\begin{equation}
d\nu=\gamma+2\beta\label{d nu beta gamma scaling law}\end{equation}
it is also well-established that the amplitude combination $\left(\frac{\xi_{0}^{+}}{\alpha_{c}}\right)^{-d}\frac{\Gamma^{+}}{B^{2}}$,
noted $\frac{R_{C}^{+}}{\left(R_{\xi}^{+}\right)^{d}}$, (using customary
notations \cite{Privman 1991}) corresponds to the universal value
of the asymptotic limit for $\Delta\tau^{*}\rightarrow0$ of the following
combination of singular properties \begin{equation}
\begin{array}{cl}
\frac{R_{C}^{+}}{\left(R_{\xi}^{+}\right)^{d}}= & 4\beta_{c}\left(\rho_{c}\right)^{2}\times\\
 & \lim\left[\frac{\kappa_{T}\left(\Delta\tau^{*}\right)\times\left[\xi\left(\Delta\tau^{*}\right)\right]^{-d}}{\left[\Delta\rho_{LV}\left(\left|\Delta\tau^{*}\right|\right)\right]^{2}}\right]_{\Delta\tau^{*}\rightarrow0^{\pm}}\end{array}\label{RCksi}\end{equation}
Equation (\ref{RCksi}) relates the singular behaviors of the correlation
length $\xi\left(\Delta\tau^{*}\right)$ {[}with critical exponent
$\nu$ and leading amplitude $\xi_{0}^{+}$] in the homogeneous domain,
the isothermal compressibility $\kappa_{T}\left(\Delta\tau^{*}\right)$
{[}with critical exponent $\gamma$ and leading amplitude $\Gamma_{0}^{+}=\frac{\Gamma^{+}}{p_{c}}$]
in the homogeneous domain and the order parameter density $\Delta\rho_{LV}\left(\left|\Delta\tau^{*}\right|\right)$
{[}with critical exponent $\beta$ and leading amplitude $B_{0}=2\rho_{c}B$]
in the non-homogeneous domain.

Using Eqs. (\ref{force balance in g field}), (\ref{Rsigmaksi}),
and (\ref{RCksi}) to eliminate both properties $\sigma\left(\left|\Delta\tau^{*}\right|\right)$
and $\Delta\rho_{LV}\left(\left|\Delta\tau^{*}\right|\right)$ of
nonhomogeneous domain, we obtain the following asymptotic equation
\begin{equation}
\begin{array}{cl}
\lim\left[S_{g}\right]_{\Delta\tau^{*}\rightarrow0^{-}}= & R_{\sigma\xi}^{+}\frac{\left(R_{C}^{+}\right)^{\frac{1}{2}}}{\left(R_{\xi}^{+}\right)^{\frac{d}{2}}}\times\frac{1}{2\left(\beta_{c}\right)^{\frac{3}{2}}\rho_{c}g}\times\\
 & \lim\left[\frac{1}{\left[\kappa_{T}\left(\Delta\tau^{*}\right)\times\xi\left(\Delta\tau^{*}\right)\right]^{\frac{1}{2}}}\right]_{\Delta\tau^{*}\rightarrow0^{+}}\end{array}\label{Sg vs khiTksi}\end{equation}
which relates the asymtotical singular behavior of the Sugden factor
in the nonhomogeneous domain to the ones of $\kappa_{T}\left(\Delta\tau^{*}\right)$
and $\xi\left(\Delta\tau^{*}\right)$ in the homogeneous domain. The
corresponding scaling law reads \begin{equation}
\left(\frac{d}{2}-1\right)\nu=\varphi-\frac{\gamma}{2}\label{d nu phi gamma scaling law}\end{equation}
The scaling laws given by Eqs. (\ref{Widom scaling law}), (\ref{d nu beta gamma scaling law}),
and (\ref{d nu phi gamma scaling law}), where explicit reference
to the space dimension is needed to connect correlation exponents
and thermodynamic exponents, are characteristic of hyperscaling and
reflect the universal features related to the two-scale-factor universality,
which do not depend on the (homogeneous or nonhomogeneous) domain
(see also Ref. \cite{single variable}).

\subsection{The master crossover of the one-component fluid subclass}

We are now able to construct one pseudo-crossover function based on
Eq. (\ref{Sg vs khiTksi}). This pseudo-crossover function for the
Sugden factor accounts exactly for the asymptotic two-scale factor
universality but agrees only qualitatively with the one-parameter
Ising-like critical crossover description at finite distance to CP.
As a matter of fact, accurate expressions of the complete classical-to-critical
crossover were recently proposed by Bagnuls and Bervillier \cite{Bagnuls 2002}
and written in appropriate Ising-like asymptotic forms by Garrabos
and Bervillier \cite{Garrabos 2006c} to account for error-bars associated
with the estimations of the universal exponents near the non-Gaussian
fixed point. Moreover, introducing only three characteristic numbers,
$\mathcal{\mathbb{L}}^{\left\{ 1f\right\} }$, $\Theta^{\left\{ 1f\right\} }$,
and $\Psi^{\left\{ 1f\right\} }$ (see Ref. \cite{Garrabos 2006e}
for details), these crossover functions can be easily modified to
accurately describe the master singular behavior of the one-component
fluid subclass. In this master description, two leading amplitudes
$\mathcal{Z}_{\chi}^{+}$, $\mathcal{Z}_{\xi}^{+}$, and one confluent
amplitude among $\mathcal{Z}_{\mathcal{\chi}}^{1,+}$ and $\mathcal{Z}_{\xi}^{1,+}$,
can be selected as characteristic parameters of the Ising-like universal
features observed in the Ising-like preasymptotic domain. $\mathcal{Z}_{\chi}^{+}$,
$\mathcal{Z}_{\xi}^{+}$, $\mathcal{Z}_{\chi}^{1,+}$, and $\mathcal{Z}_{\xi}^{1,+}$
are associated to the asymptotic crossover behavior of the correlation
length and the susceptibility in the homogeneous domain. We recall
that the corresponding master crossover functions are asymptotically
approximated by the restricted (two terms) Wegner like expansions
given by the respective equations\begin{equation}
\mathcal{\ell}^{*}\left(\mathcal{T}^{*}\right)=\mathcal{Z}_{\xi}^{+}\left(\mathcal{T}^{*}\right)^{-\nu}\left[1+\mathcal{Z}_{\xi}^{1,+}\left(\mathcal{T}^{*}\right)^{\Delta}\right]\label{two-term master elestar}\end{equation}
\begin{equation}
\mathcal{\varkappa}^{*}\left(\mathcal{T}^{*}\right)=\mathcal{Z}_{\mathcal{\chi}}^{+}\left(\mathcal{T}^{*}\right)^{-\gamma}\left[1+\mathcal{Z}_{\mathcal{\chi}}^{1,+}\left(\mathcal{T}^{*}\right)^{\Delta}\right]\label{two-term master khistar}\end{equation}
 where $\mathcal{Z}_{\mathcal{\chi}}^{+}=0.119$, $\mathcal{Z}_{\xi}^{+}=0.570$,
$\mathcal{Z}_{\chi}^{1,+}=0.555$ and $\mathcal{Z}_{\xi}^{1,+}=0.377$
are the constant values of the master (i.e. fluid independent) amplitudes,
with the universal ratio $\frac{\mathcal{Z}_{\xi}^{1,+}}{\mathcal{Z}_{\mathcal{\chi}}^{1,+}}=0.68$
\cite{Guida 1998}.%
\begin{table*}
\begin{tabular}{|c|c|c|c|c|c|c|}
\hline 
\multicolumn{1}{|c|}{(a)}&
 exponent&
 $\mathbb{Z}_{\xi}^{+}$&
 $S_{2}$&
 $i$&
 $X_{\xi,i}$&
 $Y_{\xi,i}$\tabularnewline
\hline
&
 $\nu=0.6303875$&
 $2.121008$&
 $22.9007$&
 $1$&
 $40.0606$&
 $-0.098968$\tabularnewline
\hline
&
 $\Delta=0.50189$&
&
&
 $2$&
 $11.9321$&
 $-0.15391$\tabularnewline
\hline
&
 $\nu_{MF}=0.5$&
&
&
 $3$&
 $1.90235$&
 $-0.00789505$\tabularnewline
\hline
&
&
&
&
&
 $\mathbb{Z}_{\xi}^{1,+}$&
 $5.81623$ \tabularnewline
\hline
\end{tabular}\begin{tabular}{|c|c|c|c|c|c|c|}
\hline 
\multicolumn{1}{|c|}{(b)}&
 exponent&
 $\mathbb{Z}_{\chi}^{+}$&
 $S_{2}$&
 $i$&
 $X_{\chi,i}$&
 $Y_{\chi,i}$\tabularnewline
\hline
&
 $\gamma=1.2395935$&
 $3.709601$&
 $22.9007$&
 $1$&
 $29.1778$&
 $-0.178403$\tabularnewline
\hline
&
 $\Delta=0.50189$&
&
&
 $2$&
 $11.7625$&
 $-0.282241$\tabularnewline
\hline
&
 $\gamma_{MF}=1.0$&
&
&
 $3$&
 $2.05948$&
 $-0.0185424$\tabularnewline
\hline
&
&
&
&
&
 $\mathbb{Z}_{\chi}^{1,+}$&
 $8.56347$ \tabularnewline
\hline
\end{tabular}

\caption{Values of the universal exponents and universal parameters for the
crossover functions estimated in Ref. \cite{Garrabos 2006c}; Part
(a) correlation length case in the homogeneous domain; Part (b) susceptibility
case in the homogeneous domain. \label{tab3}}
\end{table*}
 Accordingly, the modified crossover functions are given by the following
equations\begin{equation}
\begin{array}{rcl}
\frac{1}{\mathcal{\ell}^{*}\left(\mathcal{T}^{*}\right)} & = & \mathbb{Z}_{\xi}^{\left\{ 1f\right\} }\mathbb{Z}_{\xi}^{+}t^{\nu}\\
 &  & \times{\displaystyle \prod_{i=1}^{i=3}}\left[1+X_{\xi,i}t^{D\left(t\right)}\right]^{Y_{\xi,i}}\end{array}\label{elemaster complete equation}\end{equation}
\begin{equation}
\begin{array}{rcl}
\frac{1}{\mathcal{\varkappa}^{*}\left(\mathcal{T}^{*}\right)} & = & \mathbb{Z}_{\chi}^{\left\{ 1f\right\} }\mathbb{Z}_{\chi}^{+}t^{\gamma}\\
 &  & \times{\displaystyle \prod_{i=1}^{i=3}}\left[1+X_{\chi,i}t^{D\left(t\right)}\right]^{Y_{\chi,i}}\end{array}\label{khimaster complete equation}\end{equation}
 with\begin{equation}
D\left(t\right)=\frac{\Delta+\Delta_{MF}S_{2}\sqrt{t}}{1+S_{2}\sqrt{t}}\label{mean crosover exponent}\end{equation}
 and\begin{equation}
t=\Theta^{\left\{ 1f\right\} }\mathcal{T}^{*}\label{MR-master link}\end{equation}
All the critical exponents ($\nu$, $\gamma$, $\Delta$, $\Delta_{MF}$)
and the constants ($\mathbb{Z}_{\xi}^{+}$, $\mathbb{Z}_{\chi}^{+}$,
$X_{\xi,i}$, $Y_{\xi,i}$, $X_{\chi,i}$, $Y_{\chi,i}$, $S_{2}$)
of the initial crossover functions defined in Ref. \cite{Garrabos 2006c}
are reported in Table \ref{tab3}. Furthermore, in Eqs. (\ref{elemaster complete equation})
and (\ref{khimaster complete equation}), the prefactors $\mathbb{Z}_{\xi}^{\left\{ 1f\right\} }$
and $\mathbb{Z}_{\chi}^{\left\{ 1f\right\} }$ relate the asymptotic
master behavior given by Eqs. (\ref{two-term master elestar}) and
(\ref{two-term master khistar}), respectively, and satisfy to unequivocal
estimations from the three characteristic numbers $\mathcal{\mathbb{L}}^{\left\{ 1f\right\} }$,
$\Theta^{\left\{ 1f\right\} }$, and $\Psi^{\left\{ 1f\right\} }$
of the one-component fluid subclass \cite{Garrabos 2006e}, such that,

\begin{equation}
\mathbb{Z}_{\xi}^{\left\{ 1f\right\} }=\left[\mathcal{Z}_{\xi}^{+}\mathbb{Z}_{\xi}^{+}\left(\Theta^{\left\{ 1f\right\} }\right)^{\nu}\right]^{-1}\equiv\mathcal{\mathbb{L}}^{\left\{ 1f\right\} }\label{Ztab1fksi prefactor for  MR-master link}\end{equation}
 \begin{equation}
\begin{array}{rl}
\mathbb{Z}_{\chi}^{\left\{ 1f\right\} }= & \left[\mathcal{Z}_{\mathcal{X}}^{+}\mathbb{Z}_{\chi}^{+}\left(\Theta^{\left\{ 1f\right\} }\right)^{\gamma}\right]^{-1}\\
= & \left[\left(\mathcal{\mathbb{L}}^{\left\{ 1f\right\} }\right)^{d}\left(\Psi^{\left\{ 1f\right\} }\right)^{2}\right]^{-1}\end{array}\label{Ztab1fkhi prefactor for MR-master link}\end{equation}
The scale factor $\Theta^{\left\{ 1f\right\} }$ is defined from the
following ratios of the confluent amplitudes 

\begin{equation}
\Theta^{\left\{ 1f\right\} }=\left(\frac{\mathbb{\mathcal{Z}}_{\xi}^{1,+}}{\mathbb{Z}_{\xi}^{1,+}}\right)^{\frac{1}{\Delta}}=\left(\frac{\mathbb{\mathcal{Z}}_{\mathcal{X}}^{1,+}}{\mathbb{Z}_{\mathcal{X}}^{1,+}}\right)^{\frac{1}{\Delta}}\label{teta1f for MR-master link}\end{equation}
 where $\mathbb{Z}_{\xi}^{1,+}=-\sum_{i=1}^{i=3}X_{\xi,i}Y_{\xi,i}$
and $\mathbb{Z}_{\chi}^{1,+}=-\sum_{i=1}^{i=3}X_{\chi,i}Y_{\chi,i}$,
with $\frac{\mathbb{Z}_{\xi}^{1,+}}{\mathbb{Z}_{\chi}^{1,+}}=0.68$
\cite{Garrabos 2006c}. All the values of these master constants are
shown in Table \ref{tab4}. 

We also note that the \emph{}master \emph{prefactors} $\mathbb{Z}_{\xi}^{\left\{ 1f\right\} }$
and $\mathbb{Z}_{\chi}^{\left\{ 1f\right\} }$, as all the other prefactors
which modify the initial crossover functions to account for master
behavior of the renormalized properties of the one-compenent fluid
subclass, take the same value above and below the critical temperature,
while only two of them are characteristic of this subclass. In addition,
the \emph{}single master \emph{crossover parameter} $\Theta^{\left\{ 1f\right\} }$
is the same for any property along the critical isochore, above and
below the critical temperature. As demonstrated in Refs. \cite{Garrabos 2006c,Garrabos 2006e},
it is possible to define unambiguously the extension $\mathcal{T}^{*}\lesssim\mathcal{L}_{PAD}^{\left\{ 1f\right\} }$
of the preasymptotic domain where each master crossover function can
be approximated by its restricted (two-term) expansion. Using $\Theta^{\left\{ 1f\right\} }$
(see Table \ref{tab4}) we obtain\begin{equation}
\mathcal{T}^{*}\lesssim\mathcal{L}_{PAD}^{\left\{ 1f\right\} }=\frac{\mathcal{L}_{PAD}^{Ising}}{\Theta^{\left\{ 1f\right\} }}=\frac{10^{-3}}{\left(S_{2}\right)^{2}\Theta^{\left\{ 1f\right\} }}\approx5\,10^{-4}\label{PAD1f extension}\end{equation}
where $\mathcal{L}_{PAD}^{Ising}=\frac{10^{-3}}{\left(S_{2}\right)^{2}}$
is defined in Ref. \cite{Garrabos 2006c}.

\begin{table*}
\begin{tabular}{|c|c|c|}
\hline 
Correlation length&
Susceptibility&
\tabularnewline
\hline
$\nu=0.6303875$&
$\gamma=1.2396935$&
\tabularnewline
\hline
$\left(\mathbb{Z}_{\xi}^{+}\right)^{-1}=0.471474$&
$\left(\mathbb{Z}_{\chi}^{+}\right)^{-1}=0.269571$&
$\Theta^{\left\{ 1f\right\} }=4.288\,10^{-3}$\tabularnewline
\hline
$\mathcal{Z}_{\xi}^{+}=\left[\mathbb{Z}_{\xi}^{+}\mathcal{\mathbb{L}}^{\left\{ 1f\right\} }\left(\Theta^{\left\{ 1f\right\} }\right)^{\nu}\right]^{-1}=0.57$&
$\mathbb{\mathcal{Z}}_{\mathcal{\chi}}^{+}=\left[\mathbb{Z}_{\chi}^{+}\left(\mathcal{\mathbb{L}}^{\left\{ 1f\right\} }\right)^{-d}\left(\Psi^{\left\{ 1f\right\} }\right)^{-2}\left(\Theta^{\left\{ 1f\right\} }\right)^{\gamma}\right]^{-1}=0.119$&
$\Psi^{\left\{ 1f\right\} }=1.74\,10^{-4}$\tabularnewline
\hline
$\mathbb{Z}_{\xi}^{\left\{ 1f\right\} }\equiv\mathcal{\mathbb{L}}^{\left\{ 1f\right\} }=25.6988$&
$\mathbb{Z}_{\chi}^{\left\{ 1f\right\} }=\left[\left(\mathcal{\mathbb{L}}^{\left\{ 1f\right\} }\right)^{d}\left(\Psi^{\left\{ 1f\right\} }\right)^{2}\right]^{-1}=1950.7$&
$\mathcal{\mathbb{L}}^{\left\{ 1f\right\} }=25.6988$\tabularnewline
\hline
\hline 
$\Delta=0.50189$&
&
\tabularnewline
\hline
$\mathbb{Z}_{\xi}^{1,+}=0.68\mathbb{Z}_{\chi}^{1,+}=5.81623$&
$\mathbb{Z}_{\chi}^{1,+}=8.56347$&
\tabularnewline
\hline
$\mathbb{\mathcal{Z}}_{\xi}^{1,+}=\mathbb{Z}_{\xi}^{1,+}\left(\Theta^{\left\{ 1f\right\} }\right)^{\Delta}=0.68\mathbb{\mathcal{Z}}_{\mathcal{\chi}}^{1,+}=0.377$&
$\mathbb{\mathcal{Z}}_{\mathcal{\chi}}^{1,+}=\mathbb{Z}_{\chi}^{1,+}\left(\Theta^{\left\{ 1f\right\} }\right)^{\Delta}=0.555$&
\tabularnewline
\hline
&
&
\tabularnewline
\hline
\end{tabular}

\caption{Universal and master constants of Eqs. (\ref{elemaster complete equation})
and (\ref{khimaster complete equation}) for the correlation length
and the susceptibility, respectively, in the homogeneous domain (see
text and Refs. \cite{Garrabos 2006c,Garrabos 2006e} for details).
Upper part (lines 1 to 4) refers to the Ising-like leading term; The
values of the three characteristic numbers of the one component fluid
{}``subclass'' are reported in column 3, that demonstrates the unequivocal
relation between the {}``master'' crossover functions \cite{Garrabos 2006e}
and the {}``MR'' crossover functions \cite{Garrabos 2006c}. Lower
part (lines 5 to 7) refers to the first term of the confluent correction
to scaling.\label{tab4}}
\end{table*}

After appropriate rescaling of the master form of each property included
in Eq. (\ref{Sg vs khiTksi}), we define the following master quantity
\begin{equation}
\mathcal{\hat{S}}\left(\mathcal{T}^{*}\right)=R_{\sigma\xi}^{+}\frac{\left(R_{C}^{+}\right)^{\frac{1}{2}}}{\left(R_{\xi}^{+}\right)^{\frac{d}{2}}}\left[\frac{1}{\varkappa^{\ast}\left(\mathcal{T}^{*}\right)}\times\frac{1}{\ell^{\ast}\left(\mathcal{T}^{*}\right)}\right]^{\frac{1}{2}}\label{master Scalhat vs master khitstar elestar}\end{equation}
where the correlation length and the susceptibility are given by Eqs.
(\ref{elemaster complete equation}) and (\ref{khimaster complete equation}),
respectively. $\mathcal{\hat{S}}\left(\mathcal{T}^{*}\right)$ {[}Eq.
(\ref{master Scalhat vs master khitstar elestar})] is the pseudo-crossover
function of the Sugden factor which accounts for the MR description
of the classical-to-critical crossover, in the homogeneous domain
(see the discussion in next section). The corresponding curves labelled
$MR^{+}$ in Figs. \ref{fig01}c and \ref{fig01}d, confirm the perfect
agreement with the master behavior of the one-component fluid subclass
when the asymptotic term of $\mathcal{\hat{S}}\left(\mathcal{T}^{*}\right)$
corresponds to the one of $\mathcal{S}_{g^{*}}^{*}\left(\mathcal{T}^{*}\right)$
for $\mathcal{T}^{*}\rightarrow0$.

\subsection{The master leading power law of the renormalized Sugden factor}

In the preasymptotic domain defined by Eq. (\ref{PAD1f extension}),
the above formulation of the master singular behavior of $\mathcal{\hat{S}}\left(\mathcal{T}^{*}\right)$,
with $\mathcal{T}^{*}>0$, can be approximated by a restricted (two
term) expansion of equation\begin{equation}
\mathcal{\hat{S}}\left(\mathcal{T}^{*}\right)=\mathcal{Z}_{\mathcal{S}}\left(\mathcal{T}^{*}\right)^{\phi}\left[1+\mathcal{\hat{Z}}_{\mathcal{S}}^{1,+}\left(\mathcal{T}^{*}\right)^{\Delta}\right]\label{two term pseudoScalhat}\end{equation}
 where the decorated hat labels pseudo-physical quantities. Equation
(\ref{two term pseudoScalhat}) contains the asymptotic constraint
of Eq. (\ref{Sg vs khiTksi}), written following the master description
\begin{equation}
\lim\left[\mathcal{\hat{S}}\left(\mathcal{T}^{*}\right)\right]_{\mathcal{T}^{*}\rightarrow0^{+}}=\lim\left[\mathcal{S}_{g^{*}}^{*}\left(\left|\mathcal{T}^{*}\right|\right)\right]_{\mathcal{T}^{*}\rightarrow0^{-}}\label{lim pseudoScalhat vs lim S}\end{equation}
where $\mathcal{S}_{g^{*}}^{*}\left(\left|\mathcal{T}^{*}\right|\right)$,
with $\mathcal{T}^{*}<0$, is given by Eq. (\ref{two-term master Sgstar}),
while the difference occuring to the first order of the confluent
corrections to scaling is discussed below (see § 4.1). The leading
amplitude $\mathcal{Z}_{\mathcal{S}}$ has the master form \begin{equation}
\mathcal{Z}_{\mathcal{S}}=R_{\sigma\xi}^{\pm}\frac{\left(R_{C}^{+}\right)^{\frac{1}{2}}}{\left(R_{\xi}^{+}\right)^{\frac{d}{2}}}\left(\mathcal{Z}_{\chi}^{+}\mathcal{Z}_{\xi}^{+}\right)^{-\frac{1}{2}}\label{ZcalScal}\end{equation}
Using the universal values $R_{\sigma\xi}^{+}=0.376\left(\pm0.017\right)$
\cite{Gielen 1984 Ar-N2-O2-CO2-CH4,Moldover 1985,Privman 1991,Fisher 1998},
$R_{C}^{+}=0.0574\left(\pm0.0020\right)$ \cite{Bagnuls 2002}, $R_{\xi}^{+}=0.2696\left(\pm0.0007\right)$
\cite{Bagnuls 2002}, estimated for the Ising-like universality class,
and the values $\mathcal{Z}_{\chi}^{+}=0.119$, $\mathcal{Z}_{\xi}^{+}=0.57$
(see Table \ref{tab4}), we obtain\begin{equation}
\mathcal{Z}_{\mathcal{S}}=2.47\left(\pm0.17\right)\label{ZcalScal value}\end{equation}
 We note that the error-bar reported for each universal amplitude
combination only account for theoretical uncertainties on the estimated
values of the universal combinations $R_{\sigma\xi}^{+}$, $R_{C}^{+}$,
and $R_{\xi}^{+}$, while the {}``best'' central values of the master
amplitudes $\mathcal{Z}_{\chi}^{+}$ and $\mathcal{Z}_{\xi}^{+}$
are estimated using xenon as a standard one component fluid . The
large error-bar ($\pm5\%$) on $R_{\sigma\xi}^{+}$ accounts for the
theoretical values $R_{\sigma\xi}^{+}\simeq0.367\left(\pm0.009\right)$
and $R_{\sigma\xi}^{+}\simeq0.372\left(\pm0.009\right)$ estimated
by Zinn and Fisher \cite{Zinn 1996} from numerical studies of three-dimensional
Ising models, the (min and max central) values $R_{\sigma\xi}^{+}\simeq0.36\left(\pm0.01\right)$
and $R_{\sigma\xi}^{+}\simeq0.39\left(\pm0.03\right)$ quoted by Privman
et al \cite{Privman 1991} on the basis of previous theoretical calculations,
and the median values $R_{\sigma\xi}^{+}\simeq0.386\left(\pm0.1\right)$
\cite{Moldover 1985} and $R_{\sigma\xi}^{+}\simeq0.381\left(\pm0.01\right)$
\cite{Gielen 1984 Ar-N2-O2-CO2-CH4} which were initialy obtained
from the analysis of the experimental situation for fluids (see Refs.
\cite{Maass 1921 Ethane-Ethylene,Coffin 1928 i-butane,Katz 1939 2-3alcane,Stansfield 1958 Argon-Nitrogen,Smith 1967 Xenon,Grigull 1969 Carbondioxide,Gielen 1984 Ar-N2-O2-CO2-CH4,Rathjen 1977 sulfurhexafluoride-halocarbons,Straub 1980 water,Vargaftik 1983 Water,Rathjen 1980}).

The published data of the effective exponent-amplitude pair $\left\{ \varphi_{e};S_{0,e}\right\} $
reported in Table \ref{tab1} (colums 3 \& 4) allows one to validate
this leading master description at \emph{finite} distance to the critical
point, using a method equivalent to the one proposed by Moldover \cite{Moldover 1985}
to estimate $S_{0}$ by averaging the values of $\frac{S_{g}}{\left|\Delta\tau^{*}\right|^{0.935}}$
in the vicinity of $\left|\Delta\tau^{*}\right|=0.01$. The corresponding
Moldover's values (noted $S_{0,\varphi}$ to recall for the use of
the theoretical value $\varphi=0.935$), are given in column 5 of
Table \ref{tab1}. In our present work, we have estimated $S_{0,\varphi}$
by the following relation $S_{0,\varphi}=S_{0,e}\left(0.01\right)^{\varphi_{e}-0.935}$
(see also column 5, Table \ref{tab1}). From these {}``measured''
amplitude data at $\left|\Delta\tau^{*}\right|=0.01$, the corresponding
calculated values (column 6) of $\mathcal{Z}_{\mathcal{S},\varphi}=\left(\alpha_{c}\right)^{1-d}\left(g^{*}\right)^{1}\left(Z_{c}\right)^{-\frac{3}{2}}\left(Y_{c}\right)^{-\varphi}S_{0,\varphi}$
{[}see Eq. (\ref{S0 amplitude})], are in close agreement with the
asymptotic limit $\mathcal{Z}_{\mathcal{S}}=2.47$ estimated from
above hyperscaling considerations. The mean value of the data reported
in column 6 is $\left\langle \mathcal{Z}_{\mathcal{S},\varphi}\right\rangle =2.450$.
The residuals $\delta\mathcal{Z}_{\mathcal{S},\varphi}$ (column 7),
expressed in \%, are of the same order of magnitude ($\pm3.1`\%$)
than the experimental uncertainty ($\pm5`\%$) {[}see for example
the review of Moldover \cite{Moldover 1985} for a detailed analysis
of the realistic experimental errors].

This \emph{extended} master behavior is illustrated in Figs. \ref{fig01}c
and \ref{fig01}d by the curve labelled $MR_{0}$ which corresponds
to the pure power law of equation\begin{equation}
\mathcal{S}_{g^{*}}^{*}=\mathcal{Z}_{\mathcal{S}}\left|\mathcal{T}^{*}\right|^{\varphi}\label{Scalstargstar pure power law}\end{equation}
where $\mathcal{Z}_{\mathcal{S}}=2.47$ {[}see Eq. (\ref{ZcalScal value})].
In Fig. \ref{fig01}d, the two lines labelled $up$ {[}Eq. (\ref{ZcalScal value})
with $\mathcal{Z}_{\mathcal{S}}=2.64$] and $down$ {[}Eq. (\ref{ZcalScal value})
with $\mathcal{Z}_{\mathcal{S}}=2.30$], respectively, account for
the theoretical error-bar attached to this central value of $\mathcal{Z}_{\mathcal{S}}$.
Therefore, at least for a temperature-like range such that $\left|\mathcal{T}^{*}\right|<0.1$,
all the experimental results measured at finite temperature distance
to the critical point appears {}``condensed'' within these two lines.
As noted previously, such a good agreement result from the {}``universal''
median value $\varphi_{e}\equiv\varphi=0.935$ of the effective exponent
in the vicinity of $\Delta\tau^{*}=0.01$. \emph{De facto}, the asymptotical
universal features can be observed in an extended asymptotic domain,
since the confluent corrections to scaling attached to the exponent
$\Delta$ are i) only governed by the single scale factor $Y_{c}$
whatever the singular property (as already shown for the correlation
length, the susceptibility, and the order parameter density), and,
ii) certainly very small in amplitude for the Sudgen factor case.
However, the present theoretical and experimental levels of uncertainties
are of same order of magnitude and remain too high to provide an accurate
estimation of the sign and the amplitude of these (small) confluent
corrections.

As our explicit Eq. (\ref{master Scalhat vs master khitstar elestar})
is restricted only to the universal features related to hyperscaling,
there is a need for theoretical studies in the future to directly
estimate the classical-to-critical crossover of the surface tension
and the Sudgen factor in the non-homogeneous domain. Anticipating
these investigations, the following discussion gives some complementary
quantitative evaluations on the extended temperature-like range where
the asymptotic leading power law of Eq. (\ref{Scalstargstar pure power law})
can be correctly used to predict the Sudgen factor behavior (since
the applicability of the scale dilatation method goes far beyond the
one of the unvalid corresponding state principle).

\section{Discussion}

\begin{table*}
\begin{tabular}{|c|c|c|}
\hline 
Order parameter density&
Interfacial tension&
Sugden factor\tabularnewline
\hline
\hline 
$\beta=0.3257845$&
$\phi=2\nu=1.260775$&
$\varphi=\phi-\beta=0.9349905$\tabularnewline
\hline
$\mathbb{Z}_{M}=\left(R_{C}^{+}\mathcal{\mathbb{Z}}_{\chi}^{+}\right)^{-\frac{1}{2}}\left(\frac{\mathcal{\mathbb{Z}}_{\xi}^{+}}{R_{\xi}^{+}}\right)^{\frac{d}{2}}=0.937528$&
$\mathbb{Z}_{\Sigma}=R_{\sigma\xi}^{+}\left(\mathcal{\mathbb{Z}}_{\xi}^{+}\right)^{d-1}=1.6915$&
$\mathbb{Z}_{S}=R_{\sigma\xi}^{\pm}\frac{\left(R_{\xi}^{+}\right)^{\frac{d}{2}}}{\left(R_{C}^{+}\right)^{\frac{1}{2}}}\left(\mathbb{Z}_{\chi}^{+}\mathbb{Z}_{\xi}^{+}\right)^{\frac{1}{2}}=1.8042$\tabularnewline
\hline
$\mathbb{\mathcal{Z}}_{M}=\mathbb{Z}_{M}\left(\mathbb{L}^{\left\{ 1f\right\} }\right)^{d}\Psi^{\left\{ 1f\right\} }\left(\Theta^{\left\{ 1f\right\} }\right)^{\beta}=0.468$&
$\mathbb{\mathcal{Z}}_{\Sigma}=\mathbb{Z}_{\Sigma}\left(\mathbb{L}^{\left\{ 1f\right\} }\right)^{d-1}\left(\Theta^{\left\{ 1f\right\} }\right)^{\phi}=1.1558$&
$\mathbb{\mathcal{Z}}_{S}=\mathbb{Z}_{S}\frac{\left(\Theta^{\left\{ 1f\right\} }\right)^{\varphi}}{\mathbb{L}^{\left\{ 1f\right\} }\Psi^{\left\{ 1f\right\} }}=2.47$\tabularnewline
\hline
$\mathbb{Z}_{M}^{\left\{ 1f\right\} }=\left(\mathbb{L}^{\left\{ 1f\right\} }\right)^{d}\Psi^{\left\{ 1f\right\} }=2.94878$&
$\mathbb{Z}_{\Sigma}^{\left\{ 1f\right\} }\equiv\left(\mathbb{L}^{\left\{ 1f\right\} }\right)^{d-1}=660.428$&
$\mathbb{Z}_{S}^{\left\{ 1f\right\} }=\left[\mathbb{L}^{\left\{ 1f\right\} }\Psi^{\left\{ 1f\right\} }\right]^{-1}=223.634$\tabularnewline
\hline
$\frac{\left(\mathbb{Z}_{\xi}^{\left\{ 1f\right\} }\right)^{d}}{\mathbb{Z}_{\chi}^{\left\{ 1f\right\} }\left(\mathbb{Z}_{M}^{\left\{ 1f\right\} }\right)^{2}}=1$&
$\frac{\mathbb{Z}_{\Sigma}^{\left\{ 1f\right\} }}{\left(\mathbb{Z}_{\xi}^{\left\{ 1f\right\} }\right)^{d-1}}=1$&
$\frac{\mathbb{Z}_{S}^{\left\{ 1f\right\} }}{\left(\mathbb{Z}_{\xi}^{\left\{ 1f\right\} }\mathbb{Z}_{\chi}^{\left\{ 1f\right\} }\right)^{\frac{1}{2}}}=1$\tabularnewline
\hline
\hline 
$\Delta=0.50189$&
&
\tabularnewline
\hline
$\mathbb{Z}_{M}^{1}=0.9\mathbb{Z}_{\chi}^{1,+}=7.70712$&
$\frac{\mathbb{Z}_{\Sigma}^{1}}{\mathbb{Z}_{\chi}^{1,+}}$(?)&
$\frac{\mathbb{Z}_{S}^{1}}{\mathbb{Z}_{\chi}^{1,+}}$(?)\tabularnewline
\hline
$\mathbb{\mathcal{Z}}_{M}^{1}=\mathbb{Z}_{M}^{1}\left(\Theta^{\left\{ 1f\right\} }\right)^{\Delta}=0.9\mathbb{\mathcal{Z}}_{\chi}^{1,+}\approx0.5$&
$\begin{array}{rl}
\mathbb{\mathcal{Z}}_{\Sigma}^{1}\approx\mathbb{\mathcal{Z}}_{M}^{1} & \rightarrow\frac{\mathbb{\mathcal{Z}}_{\Sigma}^{1}}{\mathbb{\mathcal{Z}}_{M}^{1}}\approx1\;\;\left(see\, Eq.\,1\right)\\
\frac{\mathbb{\mathcal{Z}}_{\Sigma}^{1}}{\mathbb{\mathcal{Z}}_{\mathcal{\chi}}^{1,+}}\approx0.9 & \rightarrow\mathbb{\mathcal{Z}}_{\Sigma}^{1}\approx0.5\end{array}$&
$\begin{array}{rl}
\mathbb{\mathcal{Z}}_{S}^{1} & \approx0\;\;\left(see\, Fig.\,1\right)\\
\frac{\mathbb{\mathcal{Z}}_{S}^{1}}{\mathbb{\mathcal{Z}}_{\mathcal{\chi}}^{1,+}}=0.9\frac{\mathbb{\mathcal{Z}}_{S}^{1}}{\mathbb{\mathcal{Z}}_{M}^{1}} & \approx0\end{array}$\tabularnewline
\hline
\end{tabular}

\caption{Universal and master constants for the order parameter density (column
1), the surface tension (column 2), and the Sugden factor (column
3), in the nonhomogeneous domain. Upper part (lines 2 to 6) refers
to the Ising-like leading term; The unity value of the combinations
between the master prefactors reported in line 5 demonstrates that
the asymptotic master crossover agrees with the two-scale-factor universality
of the Ising-like systems. Lower part (lines 7 to 9) refers to the
first term of the confluent correction to scaling (see text for detail).\label{tab5}}
\end{table*}

\subsection{Ising-like universal features within the preasymptotic domain}

As demonstrated in Refs. \cite{Garrabos 2006c,Garrabos 2006e}, each
crossover function obtained from the MR scheme can be approximated
by a restricted (two term) Wegner-like expansion in the Ising-like
preasymptotic domain which extends up to\[
\left|\mathcal{T}^{*}\right|\lesssim\mathcal{L}_{PAD}^{\left\{ 1f\right\} }=\frac{\mathcal{L}_{PAD}^{Ising}}{\Theta^{\left\{ 1f\right\} }}\simeq5\,10^{-4}\]
(see the corresponding arrows in $\left|\mathcal{T}^{*}\right|$ axis
of Fig. \ref{fig01}c and \ref{fig01}d). Therefore, in addition to
Eq. (\ref{two-term master Sgstar}) related to the master singular
behavior of the renormalized Sugden factor, we are also interested
by the following similar equations\begin{equation}
\mathcal{M}_{LV}^{*}=\mathcal{Z}_{M}\left|\mathcal{T}^{*}\right|^{\beta}\left[1+\mathcal{Z}_{M}^{1}\left|\mathcal{T}^{*}\right|^{\Delta}\right]\label{two term master McalstarLV}\end{equation}
\begin{equation}
\Sigma^{*}=\mathcal{Z}_{\Sigma}\left|\mathcal{T}^{*}\right|^{\mu}\left[1+\mathcal{Z}_{\Sigma}^{1}\left|\mathcal{T}^{*}\right|^{\Delta}\right]\label{two term master sigmastar}\end{equation}
related to the master singular behaviors of the renormalized order
parameter density {[}see Eq. (\ref{master order parameter})] and
renormalized surface tension {[}see Eq. (\ref{master surface tension})],
respectively. Obviously, the hyperscaling law $d\nu=\gamma+2\beta$
provides the universal combination $\left(\mathcal{Z}_{\xi}^{+}\right)^{-d}\frac{\mathcal{Z}_{\mathcal{\chi}}^{+}}{\left(\mathcal{Z}_{M}\right)^{2}}=R_{C}^{+}\left(R_{\xi}^{+}\right)^{d}$,
while Eq. (\ref{master balance eq}) provides the {}``trivial''
relation $\mathcal{Z}_{\mathcal{S}}=\frac{\mathcal{Z}_{\Sigma}}{\mathcal{Z}_{M}}$.
Both of these amplitude combinations relate unequivocally $\mathcal{Z}_{M}$
and $\mathcal{Z}_{\Sigma}$ to the selected characteristic leading
amplitudes $\mathcal{Z}_{\chi}^{+}$ and $\mathcal{Z}_{\xi}^{+}$
of the one-component fluid subclass. Alternatively, $\mathcal{Z}_{\Sigma}$
and $\mathcal{Z}_{\xi}^{+}$ are unequivocally related by the universal
amplitude combination $R_{\sigma\xi}^{+}=\mathcal{Z}_{\Sigma}\left(\mathcal{Z}_{\xi}^{+}\right)^{d-1}$.
In such a case, we can also calculate the universal values $\mathbb{Z}_{\Sigma}=R_{\sigma\xi}^{+}\left(\mathcal{\mathbb{Z}}_{\xi}^{+}\right)^{d-1}=1.750$
and $\mathbb{Z}_{S}=\frac{\mathbb{Z}_{\Sigma}}{\mathbb{Z}_{M}}=1.867$
of the corresponding leading amplitudes for the respective crossover
functions estimated in the MR scheme {[}with $\mathcal{\mathbb{Z}}_{\xi}^{+}=2.121$,
$\mathcal{\mathbb{Z}}_{\chi}^{+}=3.7096$, and $\mathbb{Z}_{M}=\left(R_{C}^{+}\mathcal{\mathbb{Z}}_{\chi}^{+}\right)^{-\frac{1}{2}}\left(\frac{\mathcal{\mathbb{Z}}_{\xi}^{+}}{R_{\xi}^{+}}\right)^{\frac{d}{2}}=0.9375$;
see Ref. \cite{Garrabos 2006c} for detail]. Furthermore, in the relations
{[}similar to Eqs. (\ref{elemaster complete equation}) and (\ref{khimaster complete equation})]
which define the master crossover functions for the order parameter
density, the surface tension and the Sugden factor, the respective
prefactors $\mathbb{Z}_{M}^{\left\{ 1f\right\} }$, $\mathbb{Z}_{\Sigma}^{\left\{ 1f\right\} }$,
and $\mathbb{Z}_{S}^{\left\{ 1f\right\} }$ account for their unequivocal
estimation only using the three characteristic numbers $\mathbb{L}^{\left\{ 1f\right\} }$,
$\Theta^{\left\{ 1f\right\} }$, and $\Psi^{\left\{ 1f\right\} }$
of the one-component fluid subclass, such that,\begin{equation}
\begin{array}{rl}
\mathbb{Z}_{M}^{\left\{ 1f\right\} }= & \frac{\mathcal{Z}_{\mathcal{M}}}{\mathbb{Z}_{M}\left(\Theta^{\left\{ 1f\right\} }\right)^{\beta}}\\
= & \left(\mathbb{L}^{\left\{ 1f\right\} }\right)^{d}\Psi^{\left\{ 1f\right\} }\end{array}\label{Ztab1fM prefactor for MR-master link}\end{equation}

\begin{equation}
\mathbb{Z}_{\Sigma}^{\left\{ 1f\right\} }=\frac{\mathcal{Z}_{\Sigma}}{\mathbb{Z}_{\Sigma}^{+}\left(\Theta^{\left\{ 1f\right\} }\right)^{\phi}}\equiv\left(\mathbb{L}^{\left\{ 1f\right\} }\right)^{d-1}\label{Ztab1fsigma prefactor for  MR-master link}\end{equation}
 \begin{equation}
\begin{array}{rl}
\mathbb{Z}_{S}^{\left\{ 1f\right\} }= & \frac{\mathcal{Z}_{\mathcal{S}}}{\mathbb{Z}_{S}\left(\Theta^{\left\{ 1f\right\} }\right)^{\varphi}}\\
= & \left[\mathbb{L}^{\left\{ 1f\right\} }\Psi^{\left\{ 1f\right\} }\right]^{-1}\end{array}\label{Ztab1fsugden prefactor for MR-master link}\end{equation}
Equations (\ref{Ztab1fM prefactor for MR-master link}) to (\ref{Ztab1fsugden prefactor for MR-master link})
close the master representation of the singular behavior of the renormalized
interfacial properties in the nonhomogeneous domain, in agreement
with the two-scale factor universality of the Ising-like systems {[}see
the corresponding values of the universal and master quantities reported
in Table \ref{tab5}]. 

Now, using Eq. (\ref{master balance eq}) to compare the respective
first confluent amplitudes of Eqs. (\ref{two-term master Sgstar}),
(\ref{two term master sigmastar}), (\ref{two term master McalstarLV}),
we obtain $\mathcal{Z}_{\mathcal{S}}^{1}=\mathcal{Z}_{\Sigma}^{1}-\mathcal{Z}_{M}^{1}$.
>From Fig. \ref{fig01}d, the asymptotic master singular behavior expected
for $\mathcal{S}_{g^{*}}^{*}\left(\left|\mathcal{T}^{*}\right|\right)$
is compatible with the following universal values of the corresponding
amplitude ratios\begin{equation}
\begin{array}{cl}
\frac{\mathcal{Z}_{\Sigma}^{1}}{\mathcal{Z}_{M}^{1}} & \simeq1\\
\frac{\mathcal{Z}_{\mathcal{S}}^{1}}{\mathcal{Z}_{M}^{1}} & =0.9\frac{\mathcal{Z}_{\mathcal{S}}^{1}}{\mathcal{Z}_{\chi}^{1,+}}\simeq0\end{array}\label{expected universal ratio}\end{equation}
Such hypothesized {}``universal ratios'' of Eq. (\ref{expected universal ratio})
are consistent with Ising-like universal features of the asymptotic
crossover estimated from the MR scheme, which are only characterized
by a single confluent amplitude within the Ising-like preasymptotic
domain. Here these universal features are preserved via the universal
ratio value $\frac{\mathcal{Z}_{M}^{1}}{\mathcal{Z}_{\chi}^{1,+}}\simeq0.9$,
selecting $\mathcal{Z}_{\mathcal{\chi}}^{1,+}$ as a characteristic
confluent amplitude (see § 3.2 above). However, it is also important
to note that this expected crossover must satisfy the scaling law
$\varphi=\phi-\beta$ in the infinite limit $\left|\mathcal{T}^{*}\right|\rightarrow\infty$,
which leads to the mean field value $\varphi_{MF}=1$, using the mean-field
values $\beta_{MF}=\frac{1}{2}$ and $\phi_{MF}=\frac{3}{2}$ \cite{Widom 1996}.
In the range $\left|\mathcal{T}^{*}\right|>\infty$, the experimental
results reported in Fig. \ref{fig01}a to \ref{fig01}d are in disagreement
with such a mean-field prediction (see also below the § 4.3).

In addition, we note that the hyperscaling description using a pseudo-crossver
function issued from singular properties in the homogeneous domain
generates uncorrect results in the complete temperature range, i.e.
from the first-order contribution of Ising-like confluent exponent
$\Delta$ until the leading contribution related to the mean-field
exponent $\varphi_{MF}$.

For example, in our scheme based on the hyperscaling law $\varphi=\frac{\gamma+\nu}{2}$
{[}see Eq. (\ref{d nu phi gamma scaling law})], the confluent amplitude
$\mathcal{\hat{Z}}_{\mathcal{S}}^{1,+}$ in Eq. (\ref{two term pseudoScalhat})
can be made equal to $\mathcal{\hat{Z}}_{\mathcal{S}}^{1,+}=\frac{1}{2}\left(\mathcal{Z}_{\chi}^{1,+}+\mathcal{Z}_{\xi}^{1,+}\right)\approx0.466$,
leading to a universal ratio $\frac{\mathcal{\hat{Z}}_{\mathcal{S}}^{1,+}}{\mathcal{Z}_{\chi}^{1,+}}=\frac{1}{2}\left(1+\frac{\mathcal{Z}_{\xi}^{1,+}}{\mathcal{Z}_{\chi}^{1,+}}\right)\simeq0.84$
which is different from zero.

Similarly, a description based only on the hyperscaling law $\varphi=2\nu-\beta$
{[}see Eq. (\ref{Widom scaling law})] needs to replace the interfacial
tension by the inverse squared correlation length in Eq. (\ref{master balance eq}),
and provides another pseudo-crossover function, given by the equation\begin{equation}
\tilde{\mathcal{S}}\left(\left|\mathcal{T}^{*}\right|\right)=R_{\sigma\xi}^{+}\times\frac{1}{\mathcal{M}_{LV}^{*}\left(\left|\mathcal{T}^{*}\right|\right)}\times\left[\frac{1}{\ell^{\ast}\left(\mathcal{T}^{*}\right)}\right]^{2}\label{master Scaltilde vs elestar Mcalstar}\end{equation}
 where the decorated tilde distinguishs new pseudo-physical quantities
from those of Eq. (\ref{master Scalhat vs master khitstar elestar}).
In that case, a mixing occurs between properties in the homogeneous
($\ell^{\ast}\left(\mathcal{T}^{*}\right)$) and nonhomogeneous ($\mathcal{M}_{LV}^{*}\left(\left|\mathcal{T}^{*}\right|\right)$)
domains. In the Ising-like preasymptotic domain, accounting for the
relation with $\mathcal{Z}_{\mathcal{S}}=R_{\sigma\xi}^{\pm}\mathcal{Z}_{M}\left(\mathcal{Z}_{\xi}^{+}\right)^{-2}$,
Eq. (\ref{master Scaltilde vs elestar Mcalstar}) can be approximated
by \begin{equation}
\mathcal{\tilde{S}}\left(\mathcal{T}^{*}\right)=\mathcal{Z}_{\mathcal{S}}\left(\left|\mathcal{T}^{*}\right|\right)^{\phi}\left[1+\mathcal{\tilde{Z}}_{\mathcal{S}}^{1}\left(\left|\mathcal{T}^{*}\right|\right)^{\Delta}\right]\label{two term pseudoScaltilde}\end{equation}
In this latter scheme, the confluent amplitude $\mathcal{\tilde{Z}}_{\mathcal{S}}^{1}$
in Eq. (\ref{two term pseudoScaltilde}) was estimated equal to $\mathcal{\tilde{Z}}_{\mathcal{S}}^{1}=\mathcal{Z}_{M}^{1}+2\mathcal{Z}_{\xi}^{1,+}\approx1.254$,
leading to a universal ratio $\frac{\mathcal{\hat{Z}}_{\mathcal{S}}^{1}}{\mathcal{Z}_{\chi}^{1,+}}=0.9+2\frac{\mathcal{Z}_{\xi}^{1,+}}{\mathcal{Z}_{\chi}^{1,+}}\simeq2.26$
which is also significantly different from zero.

Looking now to the contribution of the leading term close to the Gaussian
fixed point, our pseudo-crossover functions estimated above does not
account for the appropriate mean-field-like description due to the
failure of the two hyperscaling laws $\varphi=\frac{\gamma+\nu}{2}$
(which gives uncorrect value $\varphi_{MF}=\frac{3}{4}$) and $\varphi=2\nu-\beta$
(which gives uncorrect value $\varphi_{MF}=\frac{1}{2}$) when we
use the corresponding mean-field values $\gamma_{MF}=1$, $\nu_{MF}=\frac{1}{2}$
and $\beta_{MF}=\frac{1}{2}$ .

\subsection{Ising-like master behavior in the extended asymptotic domain}

In spite of the absence of accurate theoretical modelling for interfacial
tension and Sugden factor along the VLE line, the MR description of
the master crossover observed for the one-component fluid subclass
can be used to provides a reasonable estimation of the renormalized
correlation length in the nonhomogeneous domain, using the following
equation\begin{equation}
\ell^{*}\left(\mathcal{T}^{*}<0\right)=\frac{\ell^{*}\left(\mathcal{T}^{*}>0\right)}{1.96}\label{elemaster Tcalstar negatif}\end{equation}
where $\ell^{*}\left(\mathcal{T}^{*}>0\right)$ of Eq. (\ref{elemaster complete equation})
is the renormalized correlation length in the homogeneous domain.
Eq. (\ref{elemaster Tcalstar negatif}) assumes that the universal
ratio $\frac{\ell^{*}\left(\mathcal{T}^{*}>0\right)}{\ell^{*}\left(\mathcal{T}^{*}<0\right)}=1.96$
is independent of the renormalized temperature like field. The result
(for $\mathcal{T}^{*}<0$) is illustrated as a $\ell^{*}\left(\mathcal{T}^{*}<0\right)$
graduation of the upper horizontal axis of Figs. \ref{fig01}c and
\ref{fig01}d. We recall that $\ell^{\ast}$ gives the best estimate
of the ratio $\frac{\xi}{\alpha_{c}}$ between the effective size
($\xi$) of the critical fluctuations and the effective size ($\alpha_{c}$)
of the attractive molecular interaction, the latter one being approximated
by the dispersion forces in Lennard-Jones-like fluids which extend
over a short range slighly greater than twice the equilibrium distance
$r_{e}$ between two interacting particles of finite hard core size
$\sigma$ (thus $\alpha_{c}\approx2r_{e}$, with $r_{e}\gtrapprox\sigma$).
Therefore, $\ell^{\ast}\left(\left|\mathcal{T}^{*}\right|=\mathcal{L}_{CIC}\right)\sim1$
in the upper axis of Figs \ref{fig01}c and \ref{fig01}d is a rough
estimate of the microscopic range of the molecular attractive interaction
between fluid particles. Such a thermal field limit corresponds to
the value $\mathcal{L}_{CIC}\approx0.15$ (here the supscript $CIC$
recall that the effective extend of the short-ranged molecular interaction
corresponds to the size of the critical interaction cell). Looking
then to the {}``Ising-like'' nature of $\mathcal{S}_{g^{*}}^{*}\left(\left|\mathcal{T}^{*}\right|\right)$,
we observe in Figs. \ref{fig01}c and \ref{fig01}d a noticeable extension
of the critical range associated to the condition $\ell^{\ast}\left(\left|\mathcal{T}^{*}\right|\right)\gtrsim3$.
Therefore, the extended asymptotic domain (labelled $EAD$) goes up
to the limit\begin{equation}
\left|\mathcal{T}^{*}\right|\lesssim\mathcal{L}^{\left\{ 1f\right\} }\approx0.03\label{EAD1f extension}\end{equation}
(see the corresponding arrow noted $\mathcal{L}^{\left\{ 1f\right\} }$
in $\left|\mathcal{T}^{*}\right|$ axis). Within $\left|\mathcal{T}^{*}\right|\lesssim\mathcal{L}^{\left\{ 1f\right\} }$,
the observed master behavior can be well-represented by $\mathcal{S}_{g^{*}}^{*}=\mathcal{Z}_{\mathcal{S}}\left|\mathcal{T}^{*}\right|^{\phi}$
{[}see Eq. (\ref{Scalstargstar pure power law})], in conformity with
the Ising-like universal features estimated from the MR scheme. We
note that such Ising-like nature of $\mathcal{S}_{g^{*}}^{*}$ in
this extended $\left|\mathcal{T}^{*}\right|$ range complements in
a self-consistent manner our previous analysis \cite{Garrabos 2002}
of the master behavior of the renormalized order parameter density
along the VLE line.

\subsection{Non-critical behavior beyond the Ising-like extended asymptotic domain}

In Ref. \cite{Garrabos 2002}, it was observed for the xenon case,
that the real crossover for the effective exponent $\beta_{e}$ appears
in the thermal field range $\left|\mathcal{D}_{CO}^{*}\right|\approx0.1-1$
where $\ell^{\ast}\left(\left|\mathcal{T}^{*}\right|\right)\gtreqqless1$
(see Fig. \ref{fig01}d). In terms of comparison between the correlation
length and the range of the microscopic intermolecular interaction,
the situation is similar to the one encountered in the homogeneous
domain for the real crossover for the effective exponent $\gamma_{e}$
\cite{Garrabos 2006a}. Indeed, when $\ell^{\ast}<1$, any MR crossover
function is not appropriate to account for the real non-universal
behavior of the one-component fluids. We recall for example that $\ell^{\ast}\cong\frac{1}{2}$
(see the limiting curve $m$ in Fig. \ref{fig01}d) corresponds to
a (non-master) microscopic arrangement where the direct correlation
distance between two interacting particles is $\approx r_{e}$ (i.e.,
$\xi\left(\Delta\tau^{*}<0\right)\approx r_{e}\gtrapprox\sigma$).
As previously noted in Ref. \cite{Garrabos 2002}, the nonhomogeneous
fluid is then made of coexisting gas and liquid which show significant
differences in the averaged quantity of particles inside the critical
interaction cell. Moreover, these differences increase approaching
the triple point temperature, since the low density gas tends to behave
as a perfect gas with one (i.e. non-interacting) particle within the
CIC volume, while the condensed liquid tends to minimize the configuration
energy of one particle by enclosing them in a particle cage made with
an increasing number (up to twelve for rare gas case) of the closest
neighboring (repulsive) particles (i.e. the mean size $d$ of the
cage is such that $r_{e}<d\approx\sigma$). For such {}``low'' and
{}``high'' local densities, cooperative density fluctuations have
no physical sense at length scale larger than $\alpha_{c}$ and the
non-universal characteristics of each fluid are only involved in the
thermodynamic properties, as clearly illustrated in Fig. \ref{fig01}d
for the Sugden factor case by the significative increasing differences
between the rescaled data for xenon and water in the range $\left|\mathcal{T}^{*}\right|>0.2$.

\section{Conclusion}

We have provided an asymptotic description of the singular behavior
of the renormalized Sugden factor (i.e. the renormalized squared capillary
length) of the one-component fluid subclass. This master crossover
behavior can be observed up to $\left|\mathcal{T}^{\ast}\right|\approx0.03$
(or $\ell^{\ast}\approx3)$ in the non-homogeneous domain, as already
noted for the renormalized order parameter density. In a future work,
we will show that this master crossover behavior can be useful to
estimate the parachor correlation along the VLE line.


\begin{thebibliography}{10}
\bibitem{particle mass}We admit that the molecular mass, $m_{\bar{p}}$,
of each constitutive fluid particle is a known quantity, in order
to infer $N$ value from the total mass ($M=Nm_{\bar{p}})$ measurement
of the amount of fluid matter filling the container of volume $V$.
\bibitem{single variable}On a general point of view, the concept
of two-scale-factor universality implies that the singular free energy
density belonging to the volume $\xi^{d}$ is a universal function
of the single variable $\frac{h}{\left|t\right|^{\frac{d\nu\delta}{\delta+1}}}$
. However, its practical formulation was currently made using the
single variable $\frac{t}{\left|m\right|^{\beta}}$ on the path $h=0$
(the critical isochore), mainly for more easy comparison with experimental
results.
\end{thebibliography}
\end{document}